\documentclass[aps,prb,twocolumn,floatfix,nofootinbib,superscriptaddress,longbibliography,nobibnotes]{revtex4-2}

\usepackage{amsmath}
\usepackage{amssymb}
\usepackage{mathtools}

\usepackage[T1]{fontenc}
\usepackage{amsfonts}
\usepackage{newtxtext}
\usepackage[varvw]{newtxmath}
\usepackage{dsfont}                 
\usepackage{bbold}                  
\usepackage[normalem]{ulem}         

\usepackage{array}
\usepackage{dcolumn}                

\usepackage{enumerate}
\usepackage[shortlabels]{enumitem}

\usepackage[usenames,dvipsnames,table]{xcolor}
\usepackage{graphicx}
\graphicspath{{figs/}} 

\usepackage{physics}                
\usepackage{csquotes}               
\usepackage{multirow}               
\usepackage{booktabs}               

\usepackage[caption=false]{subfig}
\usepackage[colorlinks=true,citecolor=blue,linkcolor=blue,filecolor=blue]{hyperref}
\usepackage[capitalize]{cleveref}

\usepackage{orcidlink}

\definecolor{JM}{rgb}{1,0.25,0.9}

\newcommand{\im}{\mathop{\mathrm{im}}}
\renewcommand{\b}[1]{{\boldsymbol{#1}}}

\renewcommand{\c}[1]{\mathcal{#1}}

\newcommand{\g}{\mathfrak{g}}
\renewcommand{\geq}{\geqslant}
\renewcommand{\leq}{\leqslant}
\newcommand{\PBC}{{\textrm{PBC}}}

\newcommand{\GSD}{{\mathrm{GSD}}}
\newcommand{\Stab}{\mathrm{Stab}}

\newcommand{\nn}{\nonumber}
\newcommand{\half}{{\textstyle{\frac{1}{2}}}}

\newcommand{\nsubg}{\vartriangleleft}

\begin{document}

\title{Lieb-Schultz-Mattis constraints for hyperbolic lattices}

\author{G. Shankar\,\orcidlink{0009-0008-9162-3962}}
\affiliation{Department of Physics, Yale University, New Haven, Connecticut 06511, USA}

\author{Joseph Maciejko\,\orcidlink{0000-0002-6946-1492}}
\affiliation{Department of Physics, University of Alberta, Edmonton, Alberta T6G 2E1, Canada}
\affiliation{Theoretical Physics Institute \& Quantum Horizons Alberta, University of Alberta, Edmonton, Alberta T6G 2E1, Canada}

\date{\today}

\begin{abstract}
The Lieb-Schultz-Mattis (LSM) theorem and its higher-dimensional extensions forbid the existence of a unique, symmetric, and gapped ground state at fractional fillings in quantum many-body systems with a conserved particle number (or spin angular momentum) and the conventional translation symmetry of Euclidean lattices. In this work, we propose a generalization of the LSM theorem to quantum many-body systems on hyperbolic lattices, i.e., regular tessellations of two-dimensional negatively curved space. By leveraging concepts from hyperbolic band theory in a many-body setting, we adapt Oshikawa's flux-threading argument to periodic hyperbolic lattices with a non-Euclidean (Fuchsian) translation symmetry and compute a lower-bound to the ground-state degeneracy as a function of filling and lattice geometry. We explore the consequences of LSM constraints for gapped phases of hyperbolic quantum matter and suggest frustrated spin models on hyperbolic analogs of the square and triangular lattices as promising platforms for realizing symmetric spin liquids in hyperbolic space.
\end{abstract}

\maketitle

\section{Introduction}

The Lieb-Schultz-Mattis (LSM) theorem~\cite{lieb1961} and its generalizations~\cite{affleck1986,oshikawa1997,affleck1988,oshikawa2000,hastings2004,hastings2005,nachtergaele2007} is a foundational result in condensed matter physics. It provides powerful constraints on the low-energy physics of a quantum many-body system without the need to solve its Hamiltonian. Specifically, for a system with unbroken lattice translation symmetry and a global $U(1)$ symmetry---the latter usually associated with particle number conservation, or spin rotations about the $z$ axis---the LSM theorem is a no-go theorem that forbids a trivial (i.e., unique, symmetric, and gapped) ground state at fractional fillings. The LSM theorem can be thought of as generalizing to a broad class of many-body systems the familiar fact that a partially filled electronic band cannot produce insulating behavior, unless interactions generate a Mott insulator. This occurs either through a spontaneous breakdown of lattice translation symmetry (enlargement of the unit cell)---typically via charge-density wave (CDW) or antiferromagnetic (AFM) ordering---or while preserving translation symmetry, but through the more exotic mechanism of topological order (fractionalized Mott insulator). The LSM theorem thus plays an important role in the search for models and materials with exotic ground states such as fractional quantum Hall states and quantum spin liquids.

Hyperbolic lattices~\cite{kollar2019,lenggenhager2022,zhang2022,chen2023b,zhang2023,chen2024,huang2024,yuan2025,xu2025} represent a new frontier in the rapidly developing field of synthetic quantum matter~\cite{grass2025}. Hyperbolic lattices are engineered structures whose connectivity emulates a discrete tiling of the hyperbolic plane, the two-dimensional (2D) space of constant negative curvature.  Besides their interest as a novel type of condensed-matter system~\cite{boettcher2022,maciejko2021,maciejko2022,cheng2022,kienzle2022,lenggenhager2023,shankar2024,mosseri2023,
lux2024,lux2023,yu2020,urwyler2022,liu2022,bzdusek2022,mosseri2022,huang2025}, hyperbolic lattices have generated interest as novel platforms for quantum error correction~\cite{breuckmann2016,jahn2021,higgott2024} and experimental tests of the holographic principle~\cite{boettcher2020,boyle2020,asaduzzaman2020,basteiro2022,dey2024}. As the single-particle physics of hyperbolic lattices is by now relatively well understood, an increasing amount of attention has been devoted lately to studying its quantum many-body physics. Although hyperbolic-lattice experiments have so far only realized noninteracting hopping models, interacting many-body systems, e.g., frustrated spin models, can in principle be engineered through qubit-photon interactions in circuit quantum electrodynamics~\cite{bienias2022}. Another promising platform to implement hyperbolic many-body physics is programmable quantum simulators based on Rydberg-atom arrays, which possess tunable interactions~\cite{ebadi2021} and already have the capacity to simulate non-Euclidean geometries~\cite{periwal2021}. Motivated by those prospects, recent theoretical studies have explored correlated phenomena on hyperbolic lattices such as CDW and AFM ordering~\cite{zhu2021,gotz2024,roy2024,leong2025b,leong2025c,gluscevich2025,leong2025}, altermagnetism~\cite{wang2026,petermann2026}, superconductivity~\cite{bashmakov2025,pavliuk2025}, the fractional quantum Hall effect~\cite{he2024,he2025,he2025b}, and quantum spin liquids~\cite{lenggenhager2025,dusel2025,mosseri2025,vidal2025}. However, several foundational issues related to quantum many-body physics on hyperbolic lattices remain to be elucidated, in particular, the fate of the LSM theorem. The LSM theorem relies essentially on lattice periodicity, i.e., translation symmetry, but hyperbolic lattices are not translation symmetric in the usual sense, as their underlying geometry is non-Euclidean. Thus, it is unclear whether or how the LSM theorem may be generalized to such systems.

In this work, we propose a generalization of the LSM theorem to hyperbolic lattices. Our approach is to combine the celebrated topological argument for the LSM theorem by Oshikawa~\cite{oshikawa2000} with ideas from hyperbolic band theory~\cite{maciejko2021,maciejko2022} to account for the non-Euclidean nature of the (Fuchsian) translation group of hyperbolic lattices. We first explore the effect of adiabatic flux threading in a periodic hyperbolic lattice as in Ref.~\cite{sun2024}, but at the level of the many-body wave function. Subsequently generalizing Oshikawa's momentum-counting procedure~\cite{oshikawa2000}, we show that fractional fillings generically imply a nontrivial ground-state degeneracy, and thus rule out a trivial ground state. We survey the consequences of those LSM constraints for various examples of hyperbolic many-body systems and identify two infinite families of hyperbolic lattices as promising platforms to realize fractionalized Mott insulators in hyperbolic space.

The rest of the paper is organized as follows. To make the paper self-contained for the reader's convenience, we begin in Sec.~\ref{sec:euclidean} with a brief recapitulation of Oshikawa's topological argument for the LSM theorem. In Sec.~\ref{sec:hyp}, we generalize this topological argument to hyperbolic lattices and derive our main results, i.e., LSM constraints on the ground-state degeneracy. Sec.~\ref{sec:gapped} discusses the phenomenology of various gapped phases of hyperbolic quantum matter and how the LSM constraints just derived apply to them. We conclude in Sec.~\ref{sec:conclusion} with a summary of our main results and outline several directions for future research.

\section{Warm-up: Lieb-Schultz-Mattis constraints for Euclidean lattices}
\label{sec:euclidean}

To establish basic concepts and set the stage for our consideration of hyperbolic lattices in Sec.~\ref{sec:hyp}, we begin by reviewing flux threading and momentum counting in Euclidean lattices, which leads to a topological ``proof'' of the LSM theorem~\cite{oshikawa2000}.

\subsection{Flux threading and momentum counting}
\label{sec:flux-euclidean}

We first summarize the flux-threading procedure heuristically before describing it mathematically, following the line of reasoning in Ref.~\cite{paramekanti2004}. We imagine coupling a many-particle system to an external $U(1)$ gauge field and threading $2\pi$ flux of this gauge field in a time-dependent manner through a noncontractible handle of a lattice with PBC. One can choose a gauge in which the time-dependent vector potential appearing in the Hamiltonian is spatially uniform, such that translation symmetry is preserved during the entire process. Thus, the initial and final states both have a well-defined (crystal) momentum. However, since momentum is not a gauge-invariant quantity, a comparison of initial and final momenta is only meaningful if they are both computed in the same gauge. For flux threading by an integer multiple of $2\pi$, the final Hamiltonian is related to the initial Hamiltonian by a large gauge transformation; applying the inverse of this gauge transformation to the final state before computing its momentum results in a meaningful, gauge-invariant momentum difference.

For simplicity, we consider a two-dimensional (2D) Euclidean square lattice with $L_1\times L_2$ unit cells and PBC in both directions, i.e., with the topology of a torus. We consider a Hamiltonian of the form $H_A=K_A+H_\text{int}$ where the nearest-neighbor kinetic term is
\begin{align}\label{KA}
K_A=-t\sum_r\sum_{j=1,2}e^{-iA_j} b_r^\dag b_{r+e_j}+\mathrm{h.c.},
\end{align}
where $e_1=\hat{x}$, $e_2=\hat{y}$ denote the primitive Bravais lattice vectors on the square lattice, $t$ the hopping amplitude, and the interaction term $H_\text{int}=H_\text{int}[n]$ depends only on the particle numbers $n_r=b_r^\dag b_r$. Here $b_r$ ($b_r^\dag$) annihilates (creates) a particle on site $r$, which can be either a hardcore boson or a spinless fermion such that $n_r=0,1$. A spin-1/2 system also falls under this description since it can be mapped to a system of hardcore bosons via $b_r=S_r^-$, $b_r^\dag=S_r^+$, and $n_r=S_r^z+\half$.

The Hamiltonian has a global $U(1)$ symmetry generated by the total particle number $N=\sum_r n_r$, which is a conserved quantity ($[N,H_A]=0$), and thus we can couple the system to a $U(1)$ gauge field $A_j$ via the Peierls substitution as indicated in Eq.~(\ref{KA}). (For a spin-$1/2$ system, that symmetry becomes the $U(1)$ symmetry of spin rotations about the $z$ axis.) We consider a flat connection, such that there is no magnetic flux piercing the surface of the torus itself but there are nontrivial holonomies
\begin{align}
\phi_j=\oint_{C_j}A,
\end{align}
corresponding to fluxes through the two handles $j=1,2$ of the torus. We choose a spatially uniform gauge, $A_j(t)=\phi_j(t)/L_j$, and ramp up the fluxes in a time-dependent manner from some initial time $t=0$ to a final time $t=T$ such that $\phi_j(0)=0$ and $\phi_j(T)=\Phi_j$. Starting from the initial ground state $\ket{\psi_0}$ at time $t=0$, the final state $\ket{\psi_0'}$ at time $t=T$ is given by $\ket{\psi_0'}=F(\Phi)\ket{\psi_0}$ where $F(\Phi)$ is the time-evolution operator,
\begin{align}\label{adiabatic}
F(\Phi)\equiv\c{T}e^{-i\int_0^T dt\,H_A(t)},
\end{align}
with $\c{T}$ denoting a time-ordered product. In the following, since $\phi_j=0$ at the initial time and $\phi_j=\Phi_j$ at the final time, we denote the initial Hamiltonian $H_A(t=0)$ by $H(0)$ and the final Hamiltonian $H_A(t=T)$ by $H(\Phi)$.

In addition to $U(1)$ symmetry, the Hamiltonian $H_A$ with a uniform gauge field $A_j$ is invariant under translations: $[T_j,H_A]=0$, $j=1,2$, where $T_j$ is the translation operator by a single lattice spacing:
\begin{align}
T_j b_r^{(\dag)}T_j^{-1}=b_{r+e_j}^{(\dag)}.
\end{align}
By translation symmetry, the initial ground state $\ket{\psi_0}$ can be chosen as an eigenstate of the translation operator:
\begin{align}\label{P0}
T_j\ket{\psi_0}=e^{iP^0_j}\ket{\psi_0},
\end{align}
where $\b{P}^0$ is the initial crystal momentum ($P^0_j\cong P^0_j+2\pi$). Furthermore, since $H_A$ is translation invariant, the time-evolution operator in Eq.~(\ref{adiabatic}) commutes with translations as well: $[T_j,F(\Phi)]=0$.

When each $\Phi_j$ is an integer multiple of the flux quantum $2\pi$, the final Hamiltonian $H(\Phi)$ is unitarily equivalent to the initial Hamiltonian $H(0)$. Indeed, since the interaction term $H_\text{int}=H_\text{int}[n]$ is gauge invariant, we can verify explicitly from Eq.~(\ref{KA}) that
\begin{align}\label{unitarity}
H(\Phi)=UH(0)U^{-1},
\end{align}
where the unitary ``twist operator''
\begin{align}\label{U}
U=\exp\left(i\sum_r\sum_{j=1,2}\frac{\Phi_j}{L_j}\Lambda_j(r) n_r\right)
\end{align}
corresponds to a large gauge transformation. Here, $\Lambda_j(r)=e_j\cdot r$ denotes the $j$th component of the site coordinate $r\in\mathbb{Z}^2$, and obeys the properties
\begin{align}\label{LambdaProperties}
\Lambda_j(r+r')&=\Lambda_j(r)+\Lambda_j(r'),\nn\\
\Lambda_j(-r)&=-\Lambda_j(r),\nn\\
\Lambda_j(e_k)&=\delta_{jk},\hspace{5mm}j,k=1,2.
\end{align}
Equation~(\ref{unitarity}) implies that for each eigenstate $\ket{\psi_n}$ of $H(0)$ with energy $E_n$, there is a corresponding eigenstate $\ket{\widetilde{\psi}_n}\equiv U\ket{\psi_n}$ of $H(\Phi)$ with the same energy. However, $U\ket{\psi_n}$ belongs to the Hilbert space of a lattice with PBC if and only if $\ket{\widetilde{\psi}_n}$, like $\ket{\psi_n}$, transforms trivially under a translation across the entire lattice: $T_j^{L_j}\ket{\widetilde{\psi}_n}\stackrel{!}{=}\ket{\widetilde{\psi}_n}$. We find
\begin{align}\label{TU}
T_j U=e^{-i\Phi_j N/L_j}UT_j,
\end{align}
and thus, applying this identity $L_j$ times,
\begin{align}
T_j^{L_j}\ket{\widetilde{\psi}_n}=e^{-i\Phi_j N}UT_j^{L_j}\ket{\psi_n}=e^{-i\Phi_j N}\ket{\widetilde{\psi}_n}.
\end{align}
Since $N$ is integral, this implies that $\ket{\widetilde{\psi}_n}$ obeys PBC if and only if $\Phi_j$ is an integer multiple of $2\pi$.

Since the final $H(\Phi)$ and initial $H(0)$ Hamiltonians differ by the gauge transformation (\ref{U}), to compute the momentum of the final state $\ket{\psi_0'}$ in the same gauge as that used for $\ket{\psi_0}$ (where $A_j=0$), we should first undo the gauge transformation before applying the translation operator as in Eq.~(\ref{P0}):
\begin{align}\label{TUInverse}
T_jU^{-1}\ket{\psi_0'}&=T_jU^{-1}F(\Phi)\ket{\psi_0}\nn\\
&=e^{i\Phi_jN/L_j}U^{-1}T_j F(\Phi)\ket{\psi_0}\nn\\
&=e^{i(P_j^0+\Phi_jN/L_j)}U^{-1}\ket{\psi_0'},
\end{align}
where we have used the inverse of Eq.~(\ref{TU}). Thus, the final momentum (strictly speaking, the momentum of $U^{-1}\ket{\psi_0'}$) is
\begin{align}\label{DeltaP}
P_j=P_j^0+\frac{\Phi_j N}{L_j},\hspace{5mm}j=1,2.
\end{align}
Defining the filling $\nu=N/V$ where $V=L_1L_2$ is the total number of unit cells in the periodic lattice, we have $P_1=P_1^0+\Phi_1\nu L_2$ and $P_2=P_2^0+\Phi_2\nu L_1$.


\subsection{Lieb-Schultz-Mattis constraints}
\label{sec:LSM}

Although the change in momentum (\ref{DeltaP}) is independent of which quantum phase of matter the system is in, its possible values lead to constraints on which phases can be realized depending on the filling $\nu$. Let us first assume the zero-flux Hamiltonian $H(0)$ has a unique, symmetric, gapped ground state $\ket{\psi_0}$ with energy $E_0$ on the torus, and thread fluxes $\Phi_1=2\pi$ and $\Phi_2=0$: then $P_2=P_2^0$ remains unchanged but $P_1=P_1^0+2\pi\nu L_2$. The change in energy under flux threading is proportional to the average current $J_j\propto \partial E_0/\partial\Phi_j$, but in an insulator the conductivity and thus the current vanish at zero temperature, such that the final state $\ket{\psi_0'}$ must have the same energy $E_0$ as the initial state $\ket{\psi_0}$. (Equivalently, we may assume that the flux threading is adiabatic and that the gap does not close~\cite{watanabe2018}, such that the system remains in its instantaneous ground state.) By Eq.~(\ref{unitarity}), this implies that $U^{-1}\ket{\psi_0'}$ is also a ground state of $H(0)$, but by assumption $\ket{\psi_0}$ was the only such ground state: thus $\ket{\psi_0}$ and $U^{-1}\ket{\psi_0'}$ are in fact the same state. Since they are the same state, they must have the same momentum: $P_1-P_1^0=2\pi\nu L_2\stackrel{!}{=}0$ (mod $2\pi)$ for all $L_2$, even or odd. Thus, such a trivial gapped ground state is only possible at integer filling $\nu\in\mathbb{Z}$. The original LSM theorem for 1D spin-$S$ chains with magnetization $m$ per unit cell~\cite{lieb1961,affleck1986,oshikawa1997} is recovered by setting $\nu=m+S$ and $L_2=1$.

If we assume the system is an insulator and the filling is commensurate but non-integer, $\nu=p/q$ with $p,q$ coprime integers, then there are at least $q$ degenerate ground states in the thermodynamic limit, corresponding to the $q$ distinct momentum sectors that can be reached e.g. by threading flux $\Phi_1=2\pi, 4\pi, \ldots,2\pi q$. If those distinct momentum states can be distinguished by local operators, the degeneracy corresponds to a spontaneously broken translation symmetry~\cite{lee1990}; otherwise, translation symmetry remains unbroken in the thermodynamic limit and the degeneracy is a consequence of topological order and fractionalization~\cite{wen1989,wen1991}. Examples of the latter include $\mathbb{Z}_2$ spin liquids~\cite{wen1991,read1991,sachdev1991,moessner2001} and fractionalized insulators~\cite{senthil2000} (with time-reversal invariance) and chiral spin liquids~\cite{kalmeyer1987,wen1989b} and fractional quantum Hall states~\cite{tsui1982,laughlin1983} (without time-reversal invariance).

\section{Lieb-Schultz-Mattis constraints for hyperbolic lattices}
\label{sec:hyp}

We now transpose the concepts reviewed in the previous section to hyperbolic lattices. After a brief review of PBC in hyperbolic lattices (Sec.~\ref{sec:PBC}), we generalize flux threading (Sec.~\ref{sec:flux-hyp}) and momentum counting (Sec.~\ref{sec:momentum-hyp} and \ref{sec:flux-div}) to infer LSM constraints for hyperbolic lattices (Sec.~\ref{sec:LSM-hyp}). We formalize the latter by deriving a general lower bound on the ground-state degeneracy (Sec.~\ref{sec:GSD-hyp}) which we illustrate through explicit numerical calculations for the $\{8,8\}$ lattice (Sec.~\ref{sec:88}). Finally, we explain how our results apply to arbitrary $\{p,q\}$ lattices (Sec.~\ref{sec:pq}).

\subsection{Periodic hyperbolic lattices}
\label{sec:PBC}

We briefly review salient facts about hyperbolic lattices and their geometry; detailed discussions can be found in Refs.~\cite{boettcher2022,maciejko2021,maciejko2022,chen2023}. A hyperbolic $\{p,q\}$ lattice is a regular tiling of the 2D hyperbolic plane by regular $p$-sided polygons such that $q$ such polygons meet at each vertex; the condition $(p-2)(q-2)>4$ must be satisfied for this tiling to live in hyperbolic space. The full symmetry group of a $\{p,q\}$ lattice is the (full) hyperbolic triangle group $\Delta(2,q,p)$ which is generated by elementary reflections and can be viewed as the space group of an infinite hyperbolic lattice~\cite{boettcher2022,chen2023}. By a mathematical result known as Fenchel's conjecture~\cite{fox1952}, the triangle group admits a finite-index, torsion-free normal subgroup $\Gamma\nsubg\Delta(2,q,p)$ which, by virtue of being torsion free, can be interpreted as the translation group of the hyperbolic lattice~\cite{maciejko2021}. The groups $\Delta(2,q,p)$ and $\Gamma$ are Fuchsian groups~\cite{Katok}, i.e., discrete subgroups of the isometry group $\mathrm{PSL}(2,\mathbb{R})$ of the hyperbolic plane, and thus we refer below to $\Gamma$ as the \emph{Fuchsian translation group}. Unlike the Abelian translation group $\mathbb{Z}^d$ of $d$-dimensional Euclidean lattices, $\Gamma$ is non-Abelian. It is isomorphic to the fundamental group $\pi_1(\Sigma_g)\cong\Gamma$ of a Riemann surface $\Sigma_g$ of genus $g\geq 2$ interpreted as a compactified Bravais unit cell. The $2g$ generators $\{\gamma_1,\ldots,\gamma_{2g}\}$ of $\Gamma$, which obey a single relation, correspond to elementary (i.e., nearest-neighbor) translations on the lattice, and to the $2g$ distinct noncontractible loops on $\Sigma_g$. The unit cells of a hyperbolic lattice can thus be labeled by group elements $\gamma\in\Gamma$, which are expressed as words in the $2g$ generators. The simplest hyperbolic lattice is the $\{4g,4g\}$ lattice, since it directly generalizes the square lattice $\{4,4\}$ considered in Sec.~\ref{sec:flux-euclidean}. We focus on this lattice below for simplicity but, as we explain later in Sec.~\ref{sec:pq}, our results easily generalize to arbitrary $\{p,q\}$ lattices.

The topological arguments of Sec.~\ref{sec:euclidean} rely on imposing PBC on the lattice, thus to generalize those arguments to hyperbolic lattices, one first needs to generalize the notion of PBC to the latter~\cite{sausset2007}. Mathematically, imposing on a hyperbolic lattice PBC that preserve its translation symmetry corresponds to choosing a finite-index normal subgroup $\Gamma_\PBC\nsubg\Gamma$, where the index $V=[\Gamma:\Gamma_\PBC]$ determines the number of unit cells of the periodic lattice (which we also refer to as a \emph{PBC cluster})~\cite{maciejko2022,lux2024}. Appropriately chosen sequences of PBC clusters of increasing size $V$ have been shown to accurately approximate at least certain properties of infinite hyperbolic lattices, such as the single-particle density of states of noninteracting tight-binding models~\cite{lux2024,lenggenhager2023}. Unlike a periodic 2D Euclidean lattice, which is topologically a torus, a periodic 2D hyperbolic lattice is surface of genus $h\geq 2$ where $h$ is related to the genus $g$ of the Fuchsian translation group by the Riemann-Hurwitz relation~\cite{FarkasKra},
\begin{align}\label{riemann-hurwitz}
h=V(g-1)+1.
\end{align}
Normal subgroups of a given finite index $V$ in $\Gamma$ can be constructed using group-theoretic methods~\cite{maciejko2022,lux2024}. For a given index $V$, there exist many such normal subgroups, which can be loosely viewed as describing periodic lattices with the same total number of sites but different ``aspect ratios''.

\subsection{Flux threading}
\label{sec:flux-hyp}

We proceed as in Sec.~\ref{sec:flux-euclidean} and consider a Hamiltonian of the form $H_A=K_A+H_\text{int}$ on a periodic $\{4g,4g\}$ hyperbolic lattice. The nearest-neighbor kinetic term is
\begin{align}\label{KA-hyp}
K_A=-t\sum_\gamma\sum_{j=1}^{2g}e^{-iA_j}b_\gamma^\dag b_{\gamma\gamma_j}+\mathrm{h.c.},
\end{align}
where $\{\gamma_1,\ldots,\gamma_{2g}\}$ denote the $2g$ generators of the Fuchsian translation group $\Gamma$, and $b_\gamma$ ($b_\gamma^\dag$) annihilates (creates) a hardcore boson or spinless fermion on site $\gamma\in\Gamma$. The interaction term $H_\text{int}=H_\text{int}[n]$ depends only on the number operators $n_\gamma=b_\gamma^\dag b_\gamma$, e.g.,
\begin{align}\label{V-hyp}
H_\text{int}=\sum_\gamma\sum_{\rho}V(\rho)n_\gamma n_{\gamma\rho}.
\end{align}
In Eqs.~(\ref{KA-hyp}) and (\ref{V-hyp}), $\gamma$ labels the sites of the lattice, while $\rho\in\Gamma$ denotes the range of the density-density interaction $V(\rho)=V(\rho^{-1})\in\mathbb{R}$.

As before, the Hamiltonian conserves the total particle number $N=\sum_\gamma n_\gamma$, thus the system has a global $U(1)$ symmetry and we can couple it to a $U(1)$ gauge field $A_j$~\cite{sun2024}. The first important difference with the Euclidean case is that, although hyperbolic lattices tessellate a 2D manifold, there are now $2g$ independent directions along which one can apply a vector potential, due to the noncommutative nature of the Fuchsian translation group~\cite{carey1998,carey2006}. As before, we choose a uniform gauge $A_j(t)=\phi_j(t)/L_j$ and ramp up the fluxes $\phi_j(t)$ from time $t=0$ to time $t=T$ such that $\phi_j(0)=0$ and $\phi_j(T)=\Phi_j$. Here, $L_j$ is a quantity as yet undefined in the hyperbolic context, but will play an important role and will be determined later in Sec.~\ref{sec:flux-div}. The time-evolution operator $F(\Phi)$ is defined as before, Eq.~(\ref{adiabatic}), and evolves the initial ground state $\ket{\psi_0}$ to some final state $\ket{\psi_0'}=F(\Phi)\ket{\psi_0}$. As before, we denote the initial and final Hamiltonians by $H(0)$ and $H(\Phi)$, respectively.

The Hamiltonian (\ref{KA-hyp}-\ref{V-hyp}) with uniform vector potential $A_j$ is invariant under (left) translations, i.e., it commutes with the translation operator $T(\eta)$,
\begin{align}
T(\eta)b_\gamma^{(\dag)}T(\eta)^{-1}=b_{\eta\gamma}^{(\dag)},
\end{align}
for any Fuchsian group element $\eta\in\Gamma$. The invariance property $T(\eta)HT(\eta)^{-1}=H$ is shown by substituting $\gamma\rightarrow\eta^{-1}\gamma$ in the sum over $\gamma$ and using the rearrangement lemma\footnote{The only difference between an infinite lattice and a periodic finite one is that one sums over the finite group $\gamma\in\Gamma/\Gamma_\PBC$ in the latter case, but the result is the same.}. Arbitrary translations are compositions of the elementary translations $T(\gamma_j)$, $j=1,\ldots,2g$.

Besides the existence of a $2g$-dimensional vector potential $A_j$, the second important difference with Euclidean lattices is that, on a periodic hyperbolic lattice, the number $2h$ of handles in the periodic surface grows with the size of the system [Eq.~(\ref{riemann-hurwitz})], and is thus greater than $2g$. The flux $\varphi_\alpha=\oint_{C_\alpha}A$ through the handle $C_\alpha$ associated with the generator $\g_\alpha\in\Gamma_\PBC$ is~\cite{sun2024}
\begin{align}\label{varphi}
\varphi_\alpha=\sum_{j=1}^{2g}\Lambda_j(\g_\alpha)\frac{\phi_j}{L_j},\hspace{5mm}\alpha=1,\ldots,2h.
\end{align}
Here, for any group element $\gamma\in\Gamma$, the function $\Lambda_j(\gamma)$ counts the number of times the generator $\gamma_j$ appears in any word representation of $\gamma$, minus the number of times the inverse generator $\gamma_j^{-1}$ appears. Physically, one should think of $\Lambda_j(\gamma)$ as the ``length'' of the translation $\gamma$ ``along direction $j$''. More formally, $\Lambda$ is a map known as the Hurewicz homomorphism~\cite{AT},
\begin{align}\label{Lambda}
\Lambda:\Gamma\rightarrow\Gamma_\text{ab}\cong\mathbb{Z}^{2g},
\end{align}
which maps the non-Abelian group $\Gamma$ to its abelianization $\Gamma_\mathrm{ab}=\Gamma/\Gamma^{(1)}$ where $\Gamma^{(1)}=[\Gamma,\Gamma]\nsubg\Gamma$ is the commutator subgroup of $\Gamma$. The factor group $\Gamma_\mathrm{ab}$ is isomorphic to the first homology group $H_1(\Sigma_g,\mathbb{Z})\cong\mathbb{Z}^{2g}$ of the compactified unit cell $\Sigma_g$. Note that, applied to the Euclidean case, $\Lambda_j(\gamma)$ reduces to the coordinate function $\Lambda_j(r)$ used in Sec.~\ref{sec:flux-euclidean}. Indeed, since it maps $\Gamma$ to an Abelian group, $\Lambda_j(\gamma)$ obeys properties analogous to Eq.~(\ref{LambdaProperties}):
\begin{align}\label{LambdaPropertiesHyp}
\Lambda_j(\gamma\gamma')&=\Lambda_j(\gamma)+\Lambda_j(\gamma'),\nn\\
\Lambda_j(\gamma^{-1})&=-\Lambda_j(\gamma),\nn\\
\Lambda_j(\gamma_k)&=\delta_{jk},\hspace{5mm}j,k=1,\ldots,2g.
\end{align}
To the difference of Euclidean lattices, not all flux configurations $\varphi_\alpha$ can be reached by working in a translationally invariant gauge $A_j=\text{const.}$, since the space of fluxes $\varphi_\alpha$ is $2h$-dimensional while the space of uniform gauge fields $A_j\propto\phi_j$ is $2g$-dimensional~\cite{sun2024}. However, to generalize the momentum counting arguments of Sec.~\ref{sec:flux-euclidean}, we need to preserve translation symmetry. Thus, we will restrict ourselves to flux configurations $\varphi_\alpha$ that can be accessed with a spatially uniform gauge. Then, the time-evolution operator commutes with Fuchsian translations: $[T(\gamma_j),F(\Phi)]=0$.

To generalize momentum counting to hyperbolic lattices, we first need to determine which gauge field configurations $\Phi_j$ yield a Hamiltonian $H(\Phi)$ that is unitarily equivalent to the zero-flux Hamiltonian $H(0)$, in a way that is consistent with PBC. Using the properties (\ref{LambdaPropertiesHyp}), we can verify by explicit computation on the Hamiltonian (\ref{KA-hyp}-\ref{V-hyp}) that $H(\Phi)=UH(0)U^{-1}$ where the twist operator $U$ is generalized to
\begin{align}\label{twist-hyp}
U=\exp\left(i\sum_\gamma\sum_{j=1}^{2g}\frac{\Phi_j}{L_j}\Lambda_j(\gamma)n_\gamma\right).
\end{align}
As before, to each eigenstate $\ket{\psi_n}$ of $H(0)$ with energy $E_n$, there corresponds an eigenstate $\ket{\widetilde{\psi}_n}=U\ket{\psi_n}$ of $H(\Phi)$ with the same energy. However, $\ket{\widetilde{\psi}_n}$ belongs to the Hilbert space of a periodic hyperbolic lattice if and only if, like $\ket{\psi_n}$, it is invariant under a translation across the whole lattice: $T(\g_\alpha)\ket{\widetilde{\psi}_n}\stackrel{!}{=}\ket{\widetilde{\psi}_n}$, $\alpha=1,\ldots,2h$. By explicit computation, we find
\begin{align}\label{MagTrans}
T(\g_\alpha)U=e^{-i\varphi_\alpha N}UT(\g_\alpha),
\end{align}
thus $\ket{\widetilde{\psi}_n}$ obeys PBC if and only if the final gauge field configuration $\phi_j(T)=\Phi_j$ is such that the flux (\ref{varphi}) through each handle is an integer multiple of $2\pi$.

\subsection{Momentum counting}
\label{sec:momentum-hyp}

For noninteracting Hamiltonians on Euclidean lattices, single-particle wave functions can be classified by their crystal momentum, a classic result known as Bloch's theorem. As exemplified in Eq.~(\ref{P0}), the notion of crystal momentum extends beyond single-particle states to many-body states, being a fundamental consequence of translation symmetry. Hyperbolic lattices are not periodic in the Euclidean sense, but obey a noncommutative translation symmetry governed by the Fuchsian group $\Gamma$. Thus, Bloch's theorem does not naively apply to them. However, recent work has shown that Bloch's theorem can be generalized to hyperbolic lattices, a framework known as hyperbolic band theory (HBT)~\cite{maciejko2021,maciejko2022}. In HBT, unitary irreducible representations (irreps) of the Fuchsian group $\Gamma$ replace the notion of crystal momentum, and are used to classify single-particle states and define single-particle energy bands.

The first observation we make here is that the generalized crystal momentum of HBT can also be used to classify \emph{many-body} states on hyperbolic lattices. In Euclidean space, the many-body crystal momentum labels the irreps of the Abelian translation group $\mathbb{Z}^2$, which are all one-dimensional. By contrast, the non-Abelian Fuchsian group $\Gamma$ admits both one-dimensional~\cite{maciejko2021} and higher-dimensional~\cite{maciejko2022,cheng2022,kienzle2022,lenggenhager2023,shankar2024} irreps. If a many-body state on a hyperbolic lattice belongs to a $\Gamma$-irrep of dimension $d>1$, that means it belongs to a degenerate multiplet of $d$ states. To derive LSM constraints from Euclidean momentum counting in Sec.~\ref{sec:LSM}, we began by assuming a unique ground state $\ket{\psi_0}$, otherwise the LSM theorem was trivially true. Likewise here, we will assume a unique ground state $\ket{\psi_0}$ before flux threading, which means that it must belong to a one-dimensional $\Gamma$-irrep characterized by a $2g$-dimensional crystal momentum $\b{P}^0$~\cite{maciejko2021}:
\begin{align}
T(\gamma_j)\ket{\psi_0}=e^{iP_j^0}\ket{\psi_0},\hspace{5mm}j=1,\ldots,2g.
\end{align}

Assuming that the $2h$ final holonomies $\varphi_\alpha$ are all integer multiples of $2\pi$, as required by PBC, the final state $\ket{\psi_0'}=F(\Phi)\ket{\psi_0}$ also has a well-defined crystal momentum, but this momentum must be computed in the same ($A_j=0$) gauge as the zero-flux Hamiltonian: i.e., we must repeat the calculation (\ref{TUInverse}) in the hyperbolic case. Similarly to Eq.~(\ref{TU}), we find
\begin{align}\label{TU-hyp}
T(\gamma_j)U=e^{-i\Phi_jN/L_j}UT(\gamma_j),
\end{align}
and thus
\begin{align}
T(\gamma_j)U^{-1}\ket{\psi_0'}=e^{i(P_j^0+\Phi_jN/L_j)}U^{-1}\ket{\psi_0'},
\end{align}
such that the final momentum is given by
\begin{align}\label{DeltaP-hyp}
P_j=P_j^0+\frac{\Phi_j N}{L_j},\hspace{5mm}j=1,\ldots,2g,
\end{align}
which reduces to Eq.~(\ref{DeltaP}) in the Euclidean (genus-1) case. We note that flux insertion results only in a shift in the Abelian crystal momentum $\b{P}$, with no change in the dimension of the Fuchsian group irrep to which the ground state belongs (here assumed to be one-dimensional). This is consistent with previous results for the single-particle case, where adiabatic flux insertion was shown to lead to a flow in the moduli space of irreps of the Fuchsian group of fixed dimension~\cite{sun2024}. There it was found that flux insertion results in the tensor product of irreps $\chi^{(\Phi)}\otimes D^{(K)}$ with $D^{(K)}$ the Fuchsian group irrep to which a single-particle (in general non-Abelian) Bloch state belongs, and $\chi^{(\Phi)}(\gamma)=\exp(-i\sum_{j=1}^{2g}\Phi_j\Lambda_j(\gamma)/L_j)$ a 1D character from flux insertion. In the present many-body case with $N$ particles, the momentum shift comes from the tensor product of $\chi^{(\Phi)}$ with itself $N$ times, $\bigotimes^N\chi^{(\Phi)}=(\chi^{(\Phi)})^N$, which gives $N$ times the momentum shift $\Phi_j/L_j$ for a single particle.

As in the Euclidean case, we will assume that the flux threading is adiabatic and that the gap to local excitations remains finite along the entire flux path for any system size. However, to derive LSM constraints from Eq.~(\ref{DeltaP-hyp}) in the hyperbolic case, we face several difficulties. In the Euclidean case, the choice of an $L_1\times L_2$ lattice allowed us to define a filling $\nu=N/V$ for a system volume (total number of unit cells) $V=L_1L_2$. Together with the quantization condition $\Phi_j\in 2\pi\mathbb{Z}$ on the inserted flux, this let us determine the lattice of allowed momentum differences $\Delta P_j\equiv P_j-P_j^0$ under flux threading for a given filling. Hyperbolic lattices are more complicated in three respects. First, geometrically, a finite hyperbolic lattice is a subregion of the Poincar\'e disk~\cite{balazs1986} that does not correspond to a simple parallelepiped spanned by two Euclidean basis vectors, thus a formula of the type $V=L_1L_2$ (or more generally, $V=|\b{L}_1\times\b{L}_2|$) does not obviously apply. Second, as reviewed in Sec.~\ref{sec:PBC}, choosing PBC that preserve Fuchsian translation symmetry is a vastly more complicated mathematical problem than choosing two basis vectors $\b{L}_1,\b{L}_2$. One must construct a finite-index normal subgroup $\Gamma_\PBC\nsubg\Gamma$ whose index $V=[\Gamma:\Gamma_\PBC]$ determines the number of unit cells. In fact, it is not even clear how to define the quantities $L_j$, $j=1,\ldots,2g$ in the hyperbolic case. Third, the flux quantization condition applies to the $2h$ holonomies $\varphi_\alpha\in 2\pi\mathbb{Z}$, but the momentum differences $\Delta P_j=\Phi_j N/L_j$ are determined by the $2g$ gauge fields $\Phi_j$. Thus, one must work backwards to infer the quantization conditions on $\Phi_j$ from those on $\varphi_\alpha$, but it is not obvious how to do this because of the dimensional mismatch.

\subsection{The flux lattice and elementary divisors}
\label{sec:flux-div}

We focus on the third issue first, i.e., how to infer from the quantization conditions
\begin{align}\label{fluxquantization}
\frac{\varphi_\alpha}{2\pi}=\sum_{j=1}^{2g}\Lambda_j(\g_\alpha)\frac{A_j}{2\pi}\in\mathbb{Z},\hspace{5mm}\alpha=1,\ldots,2h,
\end{align}
where $A_j\equiv A_j(T)=\Phi_j/L_j$ is the gauge field at the final time $t=T$, what are the allowed values of $A_j$, which control the momentum differences $\Delta P_j=NA_j$. Interpreting $\Lambda_j(\g_\alpha)$ as the $j$th component of a $2g$-dimensional vector $\b{\Lambda}(\g_\alpha)$, we consider the lattice $\Lambda_\PBC$ generated by integer linear combinations of the $2h$ vectors $\b{\Lambda}(\g_1),\ldots,\b{\Lambda}(\g_{2h})$:
\begin{align}\label{LambdaPBC}
\Lambda_\PBC\equiv\{n_1\b{\Lambda}(\g_1)+\ldots+n_{2h}\b{\Lambda}(\g_{2h}):n_1,\ldots,n_{2h}\in\mathbb{Z}\}\subset\mathbb{Z}^{2g}.
\end{align}
Since the entries of the vectors $\b{\Lambda}(\g_\alpha)$ are themselves integral, $\Lambda_\PBC$ is a subset of the standard integer lattice $\mathbb{Z}^{2g}$ that is also a lattice, i.e., it is a sublattice of $\mathbb{Z}^{2g}$. In fact, $\Lambda_\PBC=\Lambda(\Gamma_\PBC)$ is precisely the image of the normal subgroup $\Gamma_\PBC\nsubg\Gamma$ under the Hurewicz homomorphism (\ref{Lambda}). Viewing likewise $A_j$ as the $j$th component of a $2g$-dimensional gauge field (vector potential) $\b{A}$, the condition (\ref{fluxquantization}) becomes $\b{\Lambda}(\g_\alpha)\cdot\b{A}/(2\pi)\in\mathbb{Z}$, which means that the allowed values of $\b{A}/(2\pi)$ belong to the \emph{dual lattice}\footnote{The dual $\Lambda^*=\{\b{v}\}$ of a lattice $\Lambda=\{\b{u}\}$ is the set of vectors $\b{v}$ that satisfy $\b{u}\cdot\b{v}\in\mathbb{Z}$.} $\Lambda_\PBC^*$.

To construct linearly independent basis vectors for the dual lattice, we must find linearly independent basis vectors for the direct lattice (\ref{LambdaPBC}). The vectors $\b{\Lambda}(\g_\alpha)$ themselves are necessarily linearly dependent since $h>g$, thus they do not form a valid basis. In Appendix~\ref{app:rankLambda}, we show that the rank (i.e., the dimension) of $\Lambda_\PBC$ is $2g$, thus it should have $2g$ linearly independent basis vectors. Such basis vectors should be constructed as integer linear combinations of the $2h$ generating vectors $\b{\Lambda}(\g_\alpha)$ that span the same lattice $\Lambda_\PBC$, which corresponds to a $\mathrm{GL}(2h,\mathbb{Z})$ transformation. This can be achieved by computing the \emph{Smith normal form}~\cite{Cohen} $S=P\Lambda Q$ of the $2g\times 2h$ integer matrix $\Lambda_{j\alpha}\equiv\Lambda_j(\g_\alpha)$. The Smith normal form $S$ is a $2g\times 2h$ matrix of the form $S=\Bigl(\begin{array}{cc}D & \b{0}\end{array}\Bigr)$ where $D$ is a $2g\times 2g$ diagonal matrix,
\begin{align}\label{SNF}
D=\left(\begin{array}{cccc}L_1 & & & \\
& L_2 & & \\
& & \ddots & \\
& & & L_{2g}
\end{array}\right),
\end{align}
and $\b{0}$ denotes $2h-2g$ columns of zeros. The diagonal coefficients $L_1,\ldots,L_{2g}$ are positive integers called the invariant factors or \emph{elementary divisors} of $\Lambda$, and they obey\footnote{The standard mathematical notation $n\mid m$ means ``$n$ divides $m$''.} $L_{j+1}\mid L_j$ for $1\leq j<2g$, i.e., each coefficient divides the preceding one (thus $L_1\geq L_2\geq\cdots\geq L_{2g}$). The matrices $P\in\mathrm{GL}(2g,\mathbb{Z})$ and $Q\in\mathrm{GL}(2h,\mathbb{Z})$ are square, invertible integer matrices, which implies they are also unimodular, $|\det P|=|\det Q|=1$. Since $P$ is invertible, we can form the $2g\times 2h$ matrix $\Lambda Q=P^{-1}S$ whose $2h$ columns consist of linear combinations of the $2h$ columns of $\Lambda$, i.e., the $2h$ generating vectors $\b{\Lambda}(\g_\alpha)$. This matrix can be written in block form as
\begin{align}
\Lambda Q=P^{-1}S=\Bigl(\begin{array}{cc}P^{-1}D & \b{0}\end{array}\Bigr),
\end{align}
from which it is clear that there are only $2g$ linearly independent basis vectors $\b{L}_1,\ldots,\b{L}_{2g}$, corresponding to the columns of $P^{-1}D=\Bigl(\begin{array}{cccc}\b{L}_1 & \b{L}_2 & \cdots & \b{L}_{2g}\end{array}\Bigr)$. Now, $P^{-1}$ is a $\mathrm{GL}(2g,\mathbb{Z})$ transformation, thus its columns $\b{e}_1,\ldots,\b{e}_{2g}$ form a valid basis of the standard integer lattice $\mathbb{Z}^{2g}$. Writing $P^{-1}=\Bigl(\begin{array}{cccc}\b{e}_1 & \b{e}_2 & \cdots & \b{e}_{2g}\end{array}\Bigr)$, since $D$ is diagonal, the respective bases $\{\b{L}_j\}$ and $\{\b{e}_j\}$ of $\Lambda_\PBC$ and $\mathbb{Z}^{2g}$ are related by
\begin{align}\label{LambdaPBCbasis}
\b{L}_j=L_j\b{e}_j,\hspace{5mm}j=1,\ldots,2g.
\end{align}
Thus, the elementary divisors $L_j$ give the length\footnote{Note that the $\mathrm{GL}(2g,\mathbb{Z})$ transformation $P$ does not in general preserve the Euclidean length of vectors, thus if $|\b{v}|$ is the Euclidean norm of $\b{v}$, in general $|\b{e}_j|\neq 1$ and $|\b{L}_j|\neq L_j$.} of the $\Lambda_\PBC$ basis vectors $\b{L}_j$ in the coordinate system $\{\b{e}_j\}$, and provide a natural generalization of ``linear system size'' in the hyperbolic context, which will use in the formula $A_j=\Phi_j/L_j$. In other words, from the point of view of flux threading and (Abelian) momentum counting, we can view a finite hyperbolic lattice as a $2g$-dimensional, $L_1\times L_2\times\cdots\times L_{2g}$ Euclidean lattice in a (generally non-orthogonal) coordinate system set by the basis vectors $\b{e}_1,\ldots,\b{e}_{2g}$.

Finally, to determine the allowed values of the gauge field $\b{A}$, we first construct the basis $\{\b{e}_i^*\}$ dual\footnote{Note that $\b{e}_i^*$ here does not denote the complex conjugate of $\b{e}_i$ (all vectors are real), but a set of distinct (dual) lattice vectors.} to $\{\b{e}_i\}$, which obeys
\begin{align}\label{dual-lattice}
\b{e}_i\cdot\b{e}_j^*=\delta_{ij},\hspace{5mm}i,j=1,\ldots,2g.
\end{align}
Since the $\ell$th component of vector $\b{e}_i$ is given by $(\b{e}_i)_\ell=P^{-1}_{\ell i}$, we easily verify that the dual basis vectors are given by the columns of the transpose of $P$, i.e., $P^T=\Bigl(\begin{array}{cccc}\b{e}_1^* & \b{e}_2^* & \cdots & \b{e}_{2g}^*\end{array}\Bigr)$. Since the basis (\ref{LambdaPBCbasis}) spans the lattice (\ref{LambdaPBC}), the flux quantization condition $\b{\Lambda}(\g_\alpha)\cdot\b{A}/(2\pi)\in \mathbb{Z}$ is equivalent to $\b{L}_i\cdot\b{A}/(2\pi)\in\mathbb{Z}$ for all $i=1,\ldots,2g$. Expanding the gauge field on the dual basis, $\b{A}=\sum_{j=1}^{2g}x_j\b{e}_j^*$, and using Eq.~(\ref{dual-lattice}), we find that we must have $x_j=2\pi m_j/L_j$ where $m_j\in\mathbb{Z}$. Equivalently, the basis vectors of the dual lattice $\Lambda_\PBC^*$ are $\b{L}_j^*=\b{e}_j^*/L_j$, $j=1,\ldots,2g$, as can be verified directly from Eqs.~(\ref{LambdaPBCbasis}-\ref{dual-lattice}). Since the momentum difference under flux threading is $\Delta\b{P}=N\b{A}$, we thus obtain
\begin{align}\label{DeltaP-hyp2}
\frac{\Delta\b{P}}{2\pi}=N\sum_{j=1}^{2g}\frac{m_j}{L_j}\b{e}_j^*,\hspace{5mm}m_j\in\mathbb{Z},
\end{align}
which is the main result of this section. Now, a trivial momentum difference ($\Delta\b{P}=0$) corresponds to $\Delta\b{P}$ being a reciprocal lattice vector, i.e., $\Delta\b{P}/(2\pi)\in\mathbb{Z}^{2g}$. Since the standard integer lattice $\mathbb{Z}^{2g}$ is self-dual, it is also spanned by integer linear combinations of the dual basis vectors $\{\b{e}_j^*\}$. Thus, any fractional (i.e., non-integer) coefficient $Nm_j/L_j$ corresponds to a nontrivial momentum difference $\Delta\b{P}\neq 0$.

\subsection{Lieb-Schultz-Mattis constraints}
\label{sec:LSM-hyp}

To derive LSM constraints from the momentum difference (\ref{DeltaP-hyp2}), we must first express $N=\nu V$ in this equation in terms of the filling $\nu=N/V$. In the Euclidean case (Sec.~\ref{sec:euclidean}), simplifications of the type $\nu V/L_1=\nu L_2$ and $\nu V/L_2=\nu L_1$ could be carried out against the denominators $L_1,L_2$, such that consideration of odd system sizes $L_1,L_2$ led to the desired no-go theorem. To achieve something similar in the hyperbolic case, we must first determine how the total system size $V=[\Gamma:\Gamma_\PBC]$ is related to the elementary divisors $L_1,\ldots,L_{2g}$.

On the one hand, in Eq.~(\ref{IPBC}) of Appendix~\ref{app:rankLambda}, we showed that the index $I_\PBC=[\mathbb{Z}^{2g}:\Lambda_\PBC]$ of the sublattice $\Lambda_\PBC\subset\mathbb{Z}^{2g}$ was given by $I_\PBC=V/|G^{(1)}|$ where $|G^{(1)}|$ was the order of the commutator subgroup $G^{(1)}=[G,G]$ of $G=\Gamma/\Gamma_\PBC$. On the other hand, by the elementary divisor theorem~\cite{Cohen}, the quotient $\mathbb{Z}^{2g}/\Lambda_\PBC$ is given by
\begin{align}\label{ElemDivThm}
\mathbb{Z}^{2g}/\Lambda_\PBC\cong\mathbb{Z}/L_1\mathbb{Z}\oplus\mathbb{Z}/L_2\mathbb{Z}\oplus\cdots\oplus\mathbb{Z}/L_{2g}\mathbb{Z},
\end{align}
such that the index $[\mathbb{Z}^{2g}:\Lambda_\PBC]$ is given by the product of the elementary divisors:
\begin{align}
I_\PBC=L_1L_2\cdots L_{2g}.
\end{align}
Thus, we obtain
\begin{align}\label{volume-hyp}
V=|G^{(1)}|L_1L_2\cdots L_{2g},
\end{align}
which generalizes the Euclidean formula $V=L_1L_2$. Therefore $N=\nu|G^{(1)}|L_1L_2\cdots L_{2g}$, and we obtain
\begin{align}\label{DeltaP-hyp3}
\frac{\Delta\b{P}}{2\pi}=|G^{(1)}|\nu\sum_{j=1}^{2g}m_jL_1L_2\cdots\widehat{L_j}\cdots L_{2g}\b{e}_j^*,
\end{align}
where $\widehat{L_j}$ indicates that the factor $L_j$ is omitted from the product.

To proceed further, we need more information about the possible values of the elementary divisors $L_1,\ldots,L_{2g}$ as well as $|G^{(1)}|$, which both depend on the choice of normal subgroup $\Gamma_\PBC\nsubg\Gamma$, i.e., the type of periodic cluster considered. Two main cases must be distinguished according to whether the finite group $G$ of order $V$ is Abelian or non-Abelian.



\subsubsection{Abelian clusters}

If the normal subgroup $\Gamma_\PBC\nsubg\Gamma$ is such that the factor group $G=\Gamma/\Gamma_\PBC$ is Abelian, we call the corresponding periodic hyperbolic lattice an \emph{Abelian cluster}~\cite{maciejko2022}. For an Abelian cluster, the commutator subgroup $G^{(1)}$ is trivial, i.e., $|G^{(1)}|=1$, thus Eq.~(\ref{volume-hyp}) reduces to $V=L_1L_2\cdots L_{2g}$.

The simplest type of Abelian cluster is obtained whenever $V$ is a prime number, since then $G\cong\mathbb{Z}/V\mathbb{Z}$ is necessarily isomorphic to the cyclic group of order $V$. In that case, the geometry of the PBC cluster reduces to that of a 1D chain of length $V$ with PBC. Indeed, since $V=L_1\cdots L_{2g}$ in the Abelian case, the fact that $V$ is prime implies that all but one of the elementary divisors must be equal to $1$. Since $L_{j+1}\mid L_j$ in the Smith normal form (\ref{SNF}), we must have $L_1=V$ and $L_2=\cdots=L_{2g}=1$. Thus, Eq.~(\ref{ElemDivThm}) implies $\mathbb{Z}^{2g}/\Lambda_\PBC\cong\mathbb{Z}/V\mathbb{Z}$, which also confirms Eq.~(\ref{abelianizationGG1}) since $G\cong\mathbb{Z}/V\mathbb{Z}$ is Abelian. By threading flux $m_1=1$, the momentum difference in the $\b{e}_1^*$ direction is $\Delta P_1=2\pi\nu$, which recovers the 1D LSM theorem (see Sec.~\ref{sec:LSM}). In this 1D geometry, a filling $\nu=p/q$ with $p,q$ coprime rules out a trivial gapped ground state: the ground state is at least $q$-fold degenerate in the thermodynamic limit, which corresponds either to a gapless system or a spontaneously broken translation symmetry.

If the factor group $G=\Gamma/\Gamma_\PBC$ is Abelian but its order $V$ is not prime, $G$ is equal to its abelianization and thus to the quotient (\ref{ElemDivThm}),
\begin{align}
G\cong\mathbb{Z}/L_1\mathbb{Z}\oplus\mathbb{Z}/L_2\mathbb{Z}\oplus\cdots\oplus\mathbb{Z}/L_{2g}\mathbb{Z},
\end{align}
but is in general a direct sum of cyclic groups. In this case, the first $d$ elementary divisors $L_j$ are greater than one, with $1\leq d\leq 2g$, and the geometry is effectively $d$-dimensional. Momentum counting in this case is equivalent to that for a $d$-dimensional, $L_1\times L_2\times\cdots\times L_d$ Euclidean lattice, e.g., the momentum difference in the $\b{e}_1^*$ direction for flux threading with $m_1=1$ is $\Delta P_1=2\pi\nu C$ with $C=L_2\cdots L_d$~\cite{oshikawa2000b}. Though no substitute for rigorous approaches~\cite{hastings2004,hastings2005}, the topological argument~\cite{oshikawa2000,affleck1988} used in justifications of the higher-dimensional LSM theorem can be repeated here, whereby choosing a PBC cluster such that $C$ is coprime with $q$ ensures that the ground state at filling $\nu=p/q$ ($p,q$ coprime) is at least $q$-fold degenerate. When $d>1$, the effective geometry is no longer strictly one-dimensional, and intrinsic topological order is added to the list of possible mechanisms accounting for this degeneracy in the thermodynamic limit besides gaplessness and spontaneously broken translation symmetry.

\subsubsection{Non-Abelian clusters}

If the factor group $G=\Gamma/\Gamma_\PBC$ is non-Abelian, we call the corresponding periodic hyperbolic lattice a \emph{non-Abelian cluster}~\cite{maciejko2022}. In this case, the commutator subgroup $G^{(1)}$ is no longer trivial: $|G^{(1)}|>1$. The momentum difference in the $\b{e}_1^*$ direction for flux insertion with $m_1=1$ is $\Delta P_1=2\pi\nu C$ where now $C=|G^{(1)}|L_2\dots L_d$. This can be interpreted as the geometry of the PBC cluster effectively possessing a ``compact extra dimension'' of size $|G^{(1)}|$ which is insensitive to flux threading and reflects the non-Abelian nature of the translation group. For example, consider the extreme case where $G$ is a \emph{perfect group}~\cite{Robinson}, i.e., a group equal to its commutator subgroup $G=G^{(1)}$, and being in this sense maximally non-Abelian. In this case, by Eq.~(\ref{volume-hyp}), we have $L_1\cdots L_{2g}=1$ and thus every elementary divisor must be trivial, $L_1=\cdots=L_{2g}=1$. A PBC cluster of this type is analogous to a zero-dimensional ``molecule'' with $V$ sites and a high degree of internal rotational symmetry but no translational symmetry. For example, the smallest nontrivial perfect group is the alternating group $A_5$ of order 60, which is the finite subgroup of $SO(3)$ that describes the discrete rotational symmetries of the icosahedron~\cite{CoxeterPolytopes}. If $G$ is a perfect group, the coefficients $Nm_j/L_j$ in Eq.~(\ref{DeltaP-hyp2}) become $Nm_j\in\mathbb{Z}$ and have no fractional parts, thus $\Delta\b{P}=0$ for any $\nu$ and no filling constraints can be derived. Generically however, the group $G$ is not perfect even if non-Abelian, and we expect this feature to persist in the thermodynamic limit. Indeed, $\Gamma$ itself (to which the finite group $G$ can be seen as an approximation) is not perfect, given that the quotient $\Gamma/\Gamma^{(1)}\cong\mathbb{Z}^{2g}$ is nontrivial and, in fact, infinite [see Eq.~(\ref{Lambda})].

\subsection{Ground-state degeneracy}
\label{sec:GSD-hyp}

For both Abelian and non-Abelian clusters, a general lower bound for the ground-state degeneracy can be derived more formally as follows. The number of distinct ground states reached by flux threading (which is a lower bound on the actual number of ground states) corresponds to the number of distinct lattice points\footnote{We note in passing that this enumeration of degenerate ground states by counting ``fractional'' lattice points inside a unit cell bears some similarity to the study of topological degeneracy in fractional quantum Hall fluids~\cite{wen1998}.} of $\Delta\b{P}$ from Eq.~(\ref{DeltaP-hyp2}) modulo $2\pi$. Since the $2g$ directions $\{\b{e}_j^*\}$ are linearly independent, this amount to counting the number $b_j$ of distinct rational points $\{Nm_j/L_j:m_j\in\mathbb{Z}\}$ modulo $1$ along each axis $j=1,\ldots,2g$, and multiplying those numbers together. To calculate $b_j$, one needs to simplify the fraction $N/L_j=a_j/b_j$ such that $a_j,b_j$ are coprime, which is achieved by dividing the numerator and denominator by their greatest common divisor (gcd). Thus $b_j=L_j/\gcd(N,L_j)$, and the ground-state degeneracy is
\begin{align}\label{GSD}
\GSD\geq\prod_{j=1}^{2g}\frac{L_j}{\gcd(N,L_j)}.
\end{align}
Eq.~(\ref{GSD}) provides a general algorithm for deriving LSM constraints for periodic hyperbolic lattices, and is one of the main results of our work. We first summarize the algorithm in general terms:
\begin{itemize}
\item For the desired $\{p,q\}$ lattice, determine its Bravais unit cell (see Sec.~\ref{sec:pq});
\item Construct a PBC cluster with $V$ unit cells, and choose a rational\footnote{For irrational (incommensurate) fillings $\nu\notin\mathbb{Q}$, the values of $\Delta\b{P}/(2\pi)$ in Eq.~(\ref{DeltaP-hyp2}) become irrational, and thus the predicted ground-state degeneracy is infinite. This is not unreasonable since incommensurate fillings usually result in a gapless spectrum or phase separation~\cite{oshikawa2000}.} filling $\nu$ such that the total number of particles $N=\nu V$ is integer;
\item Determine the elementary divisors $L_j$ of the cluster from the Smith normal form (\ref{SNF}) and evaluate Eq.~(\ref{GSD}).
\end{itemize}
In the remaining subsections, we give explicit examples how this algorithm can be put into practice. In the two subsections immediately following, we consider a few cases where the right-hand side of Eq.~(\ref{GSD}) can be evaluated in closed form for all $\{4g,4g\}$ lattices; and in Sec.~\ref{sec:88}, we consider more general cases where Eq.~(\ref{GSD}) must be evaluated numerically, and we do this for the $\{8,8\}$ lattice. In all cases discussed below, a trivial constraint $\GSD\geq 1$ even at fractional fillings only means that a trivial ground state is not ruled out by the flux-threading argument; the argument by itself does not construct such a ground state.

\subsubsection{Abelian clusters}

The examples discussed above can be recovered as special cases of the general formula (\ref{GSD}). For $\nu=p/q$ with $p,q$ coprime, $N=pV/q\in\mathbb{Z}$ implies that $q\mid V$ (since $q\nmid p$). Consider first the case of an Abelian cluster, with $V=L_1C$ where $C=L_2\cdots L_{2g}$, and assume that $C$ is coprime with $q$~\cite{oshikawa2000}, such that $q\mid L_1$. Then $N=p(L_1/q)C$, and we have
\begin{align}\label{gcd1}
\gcd(N,L_j)&=\gcd\bigl(p(L_1/q)L_2\cdots L_{2g},L_j)=L_j,
\end{align}
for $j=2,\ldots,2g$, and thus Eq.~(\ref{GSD}) simplifies to $\GSD\geq L_1/\gcd(N,L_1)$. But
\begin{align}\label{gcd2}
\gcd(N,L_1)=\gcd(p(L_1/q)C,L_1)=(L_1/q)\gcd(pC,q),
\end{align}
and we obtain $\GSD\geq q/\gcd(pC,q)$. But since $q$ is coprime with both $p$ and $C$, we have $\gcd(pC,q)=1$ and thus $\GSD\geq q$.

If $C$ is not coprime with $q$, the derivation above does not hold, since it is then not true in general that $q\mid L_1$. However, if $q$ itself is prime, we can show generally that $q\mid L_1$ due to the properties of elementary divisors. Indeed, if $q$ is prime, applying Euclid's lemma repeatedly, $q\mid V$ implies that $q$ has to be a prime factor of at least one of the elementary divisors, i.e., $q\mid L_k$ for some $1\leq k\leq 2g$. If $k=1$, then $q\mid L_1$ is proved; otherwise, we can write
\begin{align}
L_1=\left(\prod_{j=1}^{k-1}\frac{L_j}{L_{j+1}}\right)L_k=\left(\prod_{j=1}^{k-1}\frac{L_j}{L_{j+1}}\right)\frac{L_k}{q}q,
\end{align}
and thus $q\mid L_1$, since $L_{j+1}\mid L_j$. As a result, Eqs.~(\ref{gcd1}-\ref{gcd2}) hold and $\GSD\geq q/\gcd(pC,q)$ as before. Since $q$ is now assumed to be prime, if it divides $C$, then $\gcd(pC,q)=q$ and $\GSD\geq 1$, thus no nontrivial LSM constraint can be derived in this case.

\subsubsection{Non-Abelian clusters}

Finally, we consider non-Abelian clusters. If $G=\Gamma/\Gamma_\PBC$ is a perfect group, all the elementary divisors are trivial, $L_j=1$, thus $\gcd(N,L_j)=1$. Equation~(\ref{GSD}) then implies $\GSD\geq 1$, i.e., no nontrivial LSM constraint can be derived, as concluded earlier. If $G$ is non-Abelian but not perfect, then $|G^{(1)}|<V$ and $N=(p/q)|G^{(1)}|L_1\cdots L_{2g}$. We first observe that if $q$ divides $|G^{(1)}|$, then $\gcd(N,L_j)=L_j$ for all $2g$ elementary divisors and no nontrivial LSM constraint can be derived either. Thus, we must have $q\nmid |G^{(1)}|$ to derive a nontrivial LSM constraint. Assuming $q\nmid |G^{(1)}|$, two cases can be analyzed similarly to the previous section on Abelian clusters, with identical conclusions. If $q$ is coprime with both $|G^{(1)}|$ and $C=L_2\cdots L_{2g}$, then we must have $q\mid L_1$ and $\gcd(N,L_j)=L_j$ for $j=2,\ldots,2g$, thus we obtain $\GSD\geq q/\gcd(p|G^{(1)}|C,q)$. But since $q$ is coprime with $p$, $|G^{(1)}|$, and $C$, the denominator is one and we obtain $\GSD\geq q$. If $q\nmid |G^{(1)}|$ is prime but divides $C$, we have $q\mid L_1\cdots L_{2g}$ and thus $q\mid L_1$ necessarily by the properties of elementary divisors. Then $\GSD\geq q/\gcd(p|G^{(1)}|C,q)$ as before, but since $q\mid C$, the denominator is $q$ and we obtain $\GSD\geq 1$.

\subsection{Lieb-Schultz-Mattis constraints for the $\{8,8\}$ lattice}
\label{sec:88}

\begin{figure}[t]
\includegraphics[width=\columnwidth]{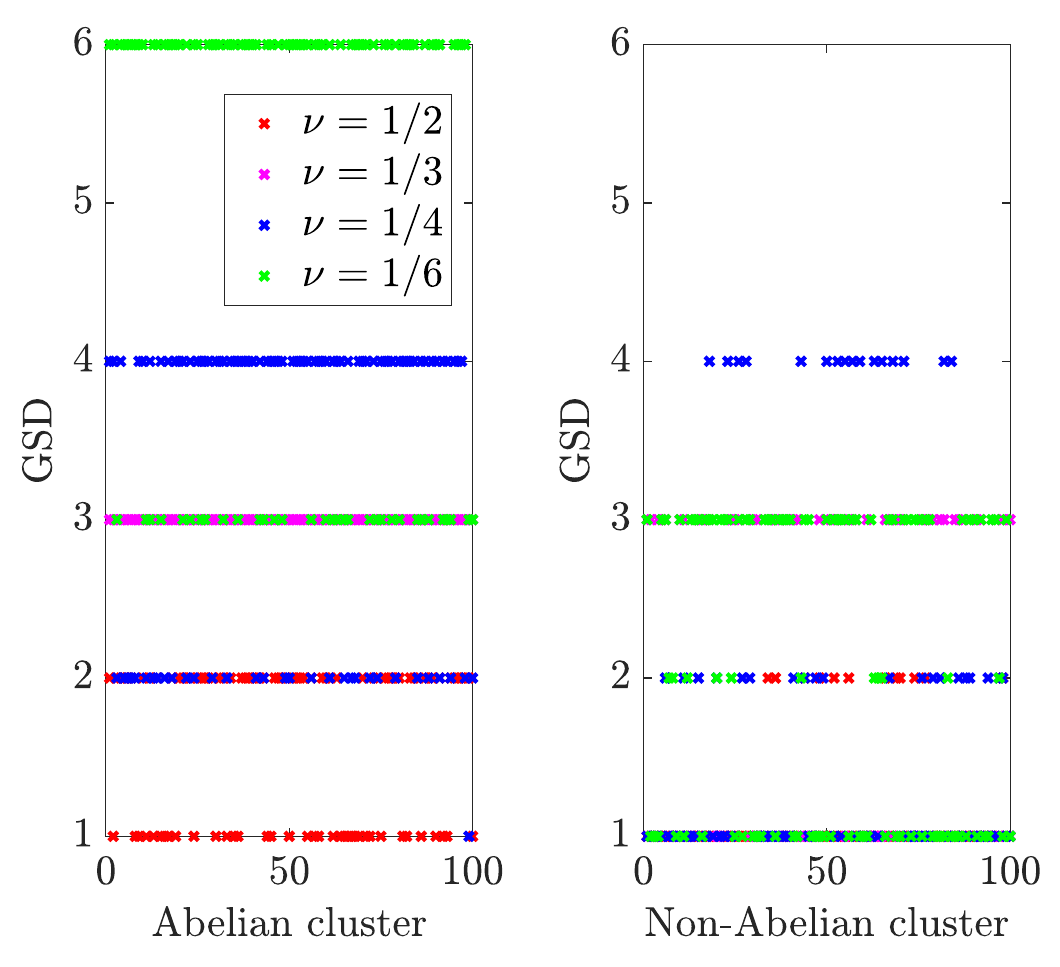}
\caption{Lower bound on the ground-state degeneracy (GSD), i.e., the right-hand side of Eq.~(\ref{GSD}), at fillings $\nu=\frac{1}{2},\frac{1}{3},\frac{1}{4},\frac{1}{6}$ for Abelian (left panel) and non-Abelian (right panel) periodic clusters of the $\{8,8\}$ lattice. Each panel shows data for 100 translation-invariant but randomly generated periodic clusters with 600 sites.}
\label{fig:GSD_2346}
\end{figure}

\begin{figure}[t]
\includegraphics[width=\columnwidth]{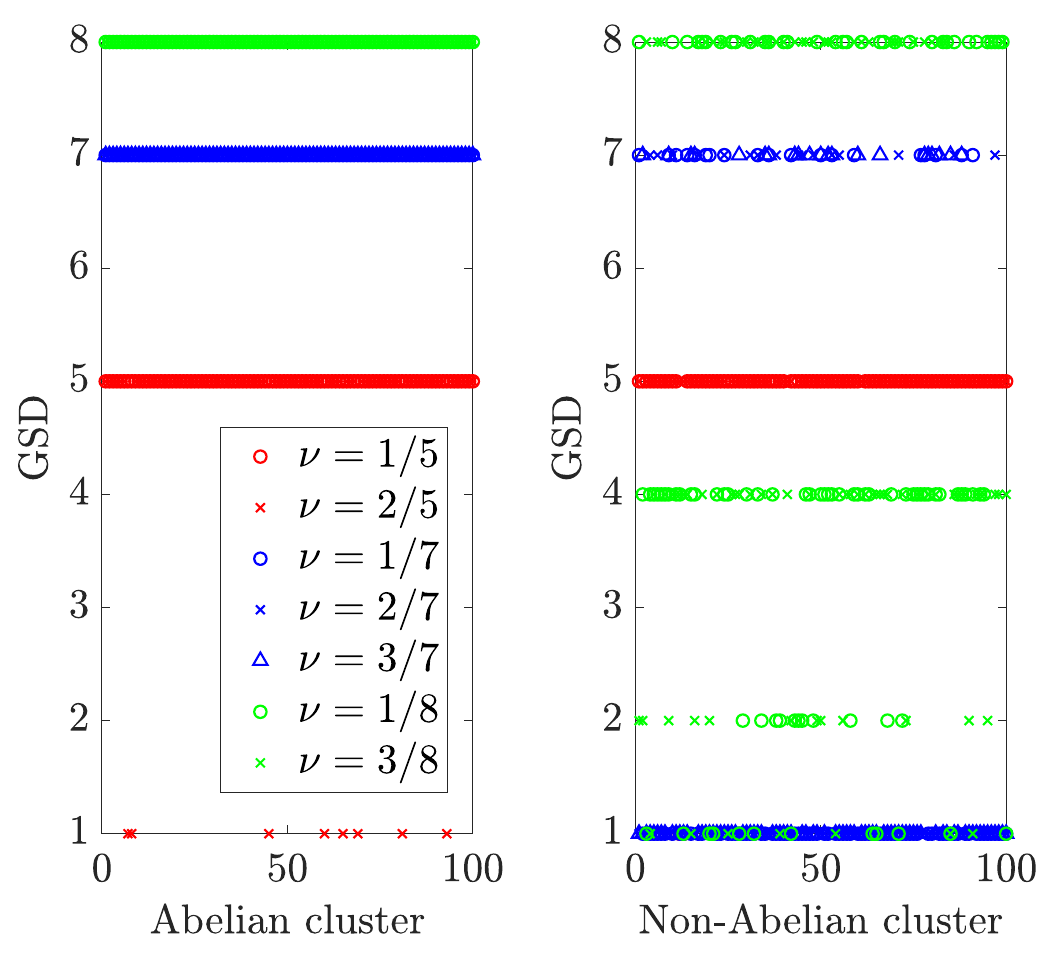}
\caption{Lower bound on the ground-state degeneracy (GSD), i.e., the right-hand side of Eq.~(\ref{GSD}), at fillings $\nu=\frac{1}{5},\frac{2}{5},\frac{1}{7},\frac{2}{7},\frac{3}{7},\frac{1}{8},\frac{3}{8}$ for Abelian (left panel) and non-Abelian (right panel) periodic clusters of the $\{8,8\}$ lattice. Each panel shows data for 100 translation-invariant but randomly generated periodic clusters with 462 sites (filings $\nu=\frac{1}{7},\frac{2}{7},\frac{3}{7}$) or 600 sites (remaining fillings).}
\label{fig:GSD_578}
\end{figure}

For general rational fillings $\nu=p/q$, the minimum ground-state degeneracy depends on the detailed geometry of the lattice and the choice of PBC. To illustrate the general formalism developed, we derive LSM constraints for the hyperbolic $\{8,8\}$ lattice by explicit numerical evaluation of the right-hand side of Eq.~(\ref{GSD}). To do this, as done in Ref.~\cite{maciejko2022}, we first use the mathematical software \textsc{GAP}~\cite{GAP4} to enumerate all normal subgroups of $\Gamma$ of index up to 25 by the low-index normal subgroups procedure~\cite{dietze1974,conder2005,FirthThesis,LINS}. We then take pairs of subgroups $H_1,H_2\nsubg\Gamma$ and compute their intersection $\Gamma_\PBC=H_1\cap H_2$ to obtain normal subgroups of larger index, and thus larger periodic lattices~\cite{tummuru2024}; we use in particular the fact that if $N_1=[\Gamma:H_1]$ and $N_2=[\Gamma:H_2]$ are coprime then $V=[\Gamma:\Gamma_\PBC]=N_1N_2$~\cite{Robinson}. The generators $\g_\alpha$ of $\Gamma_\PBC$ can be found in \textsc{GAP}, the $2g\times 2h$ matrix $\Lambda_j(\g_\alpha)$ explicitly constructed, and its Smith normal form also computed in \textsc{GAP}. Note that in practice it can be the case that $\Lambda$ has more than $2h$ columns, i.e., \textsc{GAP} returns more generators than is topologically necessary to specify the subgroup $\Gamma_\PBC$, but this only contributes more linearly dependent vectors before Smith reduction and does not change the $2g$ elementary divisors $L_j$.

We plot in Fig.~\ref{fig:GSD_2346} the right-hand side of Eq.~(\ref{GSD}) for fillings $\nu=\frac{1}{2},\frac{1}{3},\frac{1}{4},\frac{1}{6}$, and in Fig.~\ref{fig:GSD_578} for $\nu=\frac{1}{5},\frac{2}{5},\frac{1}{7},\frac{2}{7},\frac{3}{7},\frac{1}{8},\frac{3}{8}$. Note that changing $\nu\mapsto\nu+1$ does not affect the ground-state degeneracy formula since this corresponds to $N\mapsto N+V=N+|G^{(1)}|L_1\cdots L_{2g}$, but shifting $N$ by an integer multiple of $L_j$ does not change $\gcd(N,L_j)$. Likewise, the ground-state degeneracy formula is invariant under particle-hole symmetry $\nu\mapsto 1-\nu$ since this amounts to $N\mapsto V-N$, which again corresponds to a shift of $N$ by an integer multiple of $L_j$, and also $\gcd(-N,L_j)=\gcd(N,L_j)$. Thus, we can restrict ourselves to the range $0<\nu\leq\frac{1}{2}$. In both Figs.~\ref{fig:GSD_2346} and \ref{fig:GSD_578}, the left panel presents data for Abelian clusters, and the right panel data for non-Abelian clusters. We repeat the calculation for 100 clusters of each type with several hundred sites, obtained by intersection $H_1\cap H_2$ of randomly chosen pairs of normal subgroups $H_1,H_2$. Beginning with half filling $\nu=\frac{1}{2}$ in Fig.~\ref{fig:GSD_2346}, depending on the PBC cluster, the predicted degeneracy is either $\GSD\geq 2$ (nontrivial LSM constraint) or $\GSD\geq 1$ (no LSM constraint). In the 2D Euclidean case, this is akin to having a finite lattice with $L_2$ odd or even, respectively, in the formula $\Delta P_1=2\pi\nu L_2$ (see Sec.~\ref{sec:LSM}). Comparing Abelian and non-Abelian clusters, we see that a greater fraction of non-Abelian clusters do not ``see'' the nontrivial LSM constraint, which can be understood from the existence of the ``extra dimension'' $|G^{(1)}|$ in the non-Abelian case which can annihilate the LSM constraint if $|G^{(1)}|$ is even (in the half-filled case). A similar effect is seen for $\nu=\frac{1}{3}$. When $q$ in $\nu=p/q$ is not prime, as for $\nu=\frac{1}{4},\frac{1}{6}$, the minimum degeneracy is not necessarily $q$: we find that it can instead be one of the factors of $q$, and indeed in the non-Abelian case, a minimum degeneracy of 6 for $\nu=\frac{1}{6}$ is never predicted, at least for the choice of PBC clusters in Fig.~\ref{fig:GSD_2346}. Similar results are found in Fig.~\ref{fig:GSD_578}, where we also include fillings $\nu=p/q$ with $p>1$. Although for Abelian clusters, the minimum degeneracy for $\nu=p/8$ is always found to be 8, for non-Abelian clusters it can be 8, 4, 2, or 1.

\subsection{General $\{p,q\}$ lattices}
\label{sec:pq}

Finally, we comment on the applicability of our results to more general $\{p,q\}$ lattices. As mentioned in Sec.~\ref{sec:PBC}, the Fuchsian translation group $\Gamma$ of a hyperbolic lattice is a normal subgroup of its space group $\Delta\equiv\Delta(2,q,p)$, and the quotient $\c{G}=\Delta/\Gamma$ corresponds to its point group~\cite{boettcher2022}. This is a finite group whose order $|\c{G}|=[\Delta:\Gamma]$ equals the index of $\Gamma$ in $\Delta$. Geometrically, this means that a hyperbolic lattice can always be viewed as a Bravais lattice with $M$ sites per unit cell, where the precise value of $M$ depends on $p$, $q$, and the type of lattice considered (see Eq.~(\ref{MxMyMz}) in Appendix~\ref{app:wyckoff}). The $\{4g,4g\}$ lattice considered so far is a Bravais lattice with $M=1$ for any $g$~\cite{boettcher2022}, but more generally, one should generalize\footnote{An additional ``intra-cell'' term $\propto\sum_\gamma\sum_{IJ}t_{IJ}^0 b_{\gamma I}^\dag b_{\gamma J}$ should also be included in $K_A$, but is invariant under the large gauge transformation (\ref{twist-pq}) and thus does not affect the conclusions.} the Hamiltonian (\ref{KA-hyp}-\ref{V-hyp}) to
\begin{align}
K_A&=-\sum_\gamma\sum_{j=1}^{2g}\sum_{I,J=1}^M t_{IJ}(\gamma_j)e^{-iA_j}b_{\gamma I}^\dag b_{\gamma\gamma_j,J},\label{KAmultiorbital}\\
H_\text{int}&=\sum_\gamma\sum_\rho\sum_{I,J=1}^M V_{IJ}(\rho)n_{\gamma I}n_{\gamma\rho,J},
\end{align}
where the annihilation (creation) operator $b_{\gamma I}^{(\dag)}$ now features an additional ``orbital'' or sublattice index $I=1,\ldots,M$, and the hopping amplitude $t$ and interaction potential $V(\rho)$ become $M\times M$ matrices: $t_{IJ}(\gamma_j^{-1})=t_{JI}^*(\gamma_j)$ and $V_{IJ}(\rho^{-1})=V_{JI}(\rho)$. By generalizing the twist operator (\ref{twist-hyp}) to include a sum over sublattices,
\begin{align}\label{twist-pq}
U=\exp\left(i\sum_\gamma\sum_{j=1}^{2g}\sum_{I=1}^M\frac{\Phi_j}{L_j}\Lambda_j(\gamma)n_{\gamma I}\right),
\end{align}
where $n_{\gamma I}=b_{\gamma I}^\dag b_{\gamma I}$, one straightforwardly verifies that Eq.~(\ref{TU-hyp}) and thus all our previous results remain valid, except that the total particle number now includes a sum over sublattices, 
\begin{align}
N=\sum_\gamma\sum_{I=1}^M n_{\gamma I}.
\end{align}
In the expression $\nu=N/V$ for the filling, the quantity $V=[\Gamma:\Gamma_\PBC]$ is the number of Bravais unit cells, i.e., fundamental domains of the translation group $\Gamma$. Thus, in general, the filling $\nu$ to be used in the LSM constraints of Sec.~\ref{sec:LSM-hyp} corresponds to the number of particles \emph{per Bravais unit cell}, not per site of the $\{p,q\}$ lattice. In the next section, we exploit this observation to constrain the possible ground states of quantum many-body systems on various $\{p,q\}$ lattices.

\section{Gapped phases of hyperbolic quantum matter}
\label{sec:gapped}

Our main results Eq.~(\ref{DeltaP-hyp2}) [or equivalently Eq.~(\ref{DeltaP-hyp3})] and Eq.~(\ref{GSD}) establish that at integer filling $\nu\in\mathbb{Z}$ the momentum difference $\Delta\b{P}$ is always trivial, thus a trivial (unique, symmetric, gapped) ground state is possible at such fillings. By contrast, when $\nu$ is rational but non-integer, the analysis of Sec.~\ref{sec:hyp} shows that PBC cluster geometries---including non-Abelian ones, which are important to approximate the thermodynamic limit---exist such that the ground state must be degenerate. Thus, our analysis suggests that the overall statement of the Euclidean LSM theorem can be extended to hyperbolic lattices: namely that a gapped insulating phase can be trivial at integer filling but must be nontrivial at non-integer fillings. Despite the formally identical statement, a difference between Euclidean and non-Euclidean lattices persists and is reflected in the actual value of the filling as the number of particles per Bravais unit cell (see Sec.~\ref{sec:pq}), since the Bravais unit cell of a hyperbolic lattice is a nontrivial function of its geometry~\cite{boettcher2022,chen2023}. Below, we explore the implications of this statement for various examples of gapped phases of hyperbolic quantum matter.

\subsection{Trivial insulators}

The simplest example of trivial ground state is a band insulator of noninteracting fermions. Consider first the bipartite $\{8,3\}$ lattice (see, e.g., Fig.~\ref{fig:VBS}, with the two sublattices indicated by black and white dots). The nearest-neighbor hopping model on this lattice is gapless~\cite{mosseri2023}, but a staggered on-site potential with opposite sign on the two sublattices opens a gap when the entire lattice is half filled~\cite{urwyler2022}. However, as discussed in Sec.~\ref{sec:pq}, the filling $\nu$ to be used in the LSM constraints of Sec.~\ref{sec:LSM-hyp} is the number of particles per Bravais unit cell. For the $\{8,3\}$ lattice, half filling corresponds to $\frac{1}{2}$ times 16 sites per unit cell (see Appendix~\ref{app:wyckoff}), thus an integer filling $\nu=8$. This is consistent with our LSM constraints, since the staggered potential preserves the Fuchsian translation symmetry. In the presence of complex next-nearest-neighbor hopping terms that also preserve the Fuchsian translation symmetry but break time-reversal symmetry, the gap at filling $8/16$ persists but additional, topologically nontrivial gaps open at fillings $5/16$ and $11/16$, corresponding to hyperbolic analogs of Chern insulators~\cite{urwyler2022,chen2023}. Once again, those correspond to integer fillings $\nu=5$ and $\nu=11$ per Bravais unit cell, for which the absence of LSM constraint allows a band insulating ground state.

Recent studies have investigated hyperbolic tight-binding models with exactly flat bands~\cite{bzdusek2022,mosseri2022}. For both lattices, band touchings between the flat band and dispersive bands were found, such that the octagon-kagome lattice at filling $1/3$ (resp. the octagon-dice lattice at fillings $3/11$ and\footnote{More precisely, it was argued that adding a sublattice potential that preserves the Fuchsian translation symmetry can gap out one of the two band touchings, but not both~\cite{bzdusek2022}.} $8/11$) were suggested to be gapless~\cite{bzdusek2022}. However, since the number of sites per Bravais unit cell for those lattices is 24 (resp. 22, see Appendix~\ref{app:wyckoff}), the corresponding fillings are $\nu=8$ (resp. $\nu=6$ and $\nu=16$), implying that according to our LSM constraints, a unique gapped ground state can exist at those fillings. Interestingly, it was recently found that the gaplessness predicted for those lattices is in fact a finite-size artefact~\cite{lenggenhager2023,looser2026}, which is consistent with the LSM perspective presented here. For the heptagon-kagome (resp. heptagon-dice) lattices based on the $\{7,3\}$ lattice, fully gapped ground states were found at fillings $1/3$ (resp.  $3/10$ and $7/10$), with no band touching~\cite{bzdusek2022}. The point group $\c{G}=\Delta(2,3,7)/\Gamma$ of the $\{7,3\}$ lattice has 336 elements, implying 84 sites per unit cell for the heptagon-kagome lattice (resp. 80 sites per unit cell for the heptagon-dice lattice). Thus, those gapped ground states have integer fillings $\nu=28$ (resp. $\nu=24$ and $\nu=56$), also consistent with our LSM constraints.

For a less trivial example, we consider the Kitaev model on the $\{7,3\}$ lattice studied in Ref.~\cite{mosseri2025}. This is a spin-1/2 model with a gapped, fractionalized chiral spin liquid ground state that does not have any $U(1)$ spin rotation symmetry. However, it is exactly solvable by mapping to free Majorana fermions provided a 3-edge coloring of the lattice can be found. By explicit computational search, we find a 3-edge coloring of the $\{7,3\}$ lattice that is compatible with the Fuchsian translation symmetry, i.e., a 3-edge coloring of the compactified Bravais unit cell (the Klein quartic~\cite{boettcher2022}). Furthermore, the chiral spin liquid ground state found in Ref.~\cite{mosseri2025} corresponds to a homogeneous flux per heptagonal plaquette of $w_p=e^{i\pi /2}$ (or $w_p=e^{-i\pi /2}$). There are 24 such heptagonal plaquettes within the unit cell~\cite{EightfoldWay}, thus the net flux per unit cell is trivial, $\prod_{p=1}^{24}w_p=1$, and a gauge that preserves Fuchsian translation symmetry can be chosen. As a result, the free Majorana Hamiltonian can be written as
\begin{align}\label{KSL}
H=\frac{i}{4}\sum_\gamma\sum_{j=1}^{2g}\sum_{I,J=1}^MA_{IJ}(\gamma_j)c_{\gamma I}c_{\gamma\gamma_j,J},
\end{align}
where the $c$ are Majorana operators, $g=3$ is the genus of the compactified unit cell, $M=(336/2)/3=56$ is the number of sites per unit cell [see Eq.~(\ref{MxMyMz})], and $A_{IJ}(\gamma_j^{-1})=-A_{JI}(\gamma_j)$ is a real antisymmetric matrix. Although there is no physical $U(1)$ symmetry, the (particle-hole symmetric) single-particle spectrum of the Majorana Hamiltonian (\ref{KSL}) is the same as that of a tight-binding model (\ref{KAmultiorbital}) of complex spinless fermions\footnote{More formally, the Hamiltonian (\ref{KSL}) can be written in Bogoliubov-de Gennes form with two-component Nambu spinors $\psi,\psi^\dag$ and a fictitious $U(1)$ symmetry under $\psi\mapsto e^{i\theta}\psi$, $\psi^\dag\mapsto e^{-i\theta}\psi^\dag$.} with Hermitian hopping matrix $t_{IJ}(\gamma_j)=iA_{IJ}(\gamma_j)$. The corresponding single-particle density of states was computed in Ref.~\cite{mosseri2025} using exact diagonalization in real space. Six ``energy bands'' separated by energy gaps were found: two bands with spectral weight $1/14$ and four bands with spectral weight $3/14$, corresponding to ``band insulating'' ground states at fillings $1/14$, $4/14=2/7$, $7/14=1/2$, $10/14=5/7$, and $13/14$. Indeed, those respectively correspond to integer fillings $\nu=4,16,28,40,52$ per unit cell, for which our LSM constraints allow a trivial gapped ground state.

We point out in this context that unusual features of hyperbolic tight-binding models like spectral weights being different in each ``band''~\cite{mosseri2025} or gaps at seemingly ``fractional'' fillings---like in the Chern insulating, flat-band, and Majorana hopping models discussed above---are easily understood from the perspective of Fuchsian translation symmetry and hyperbolic band theory. The Bravais unit cell contains $M$ sites (Appendix~\ref{app:wyckoff}) and thus one expects $M$ Bloch bands in hyperbolic reciprocal space~\cite{maciejko2021,maciejko2022,sun2024}. An isolated region of nonzero density of states generically corresponds to an integer number $B\in\{1,\ldots,M\}$ of overlapping Bloch bands and thus carries net spectral weight $B/M$.

\subsection{Broken translational symmetry}

\begin{table}[t!]
\begin{ruledtabular}
\begin{tabular}{cll}
lattice	&   $|\c{G}^+|$    &	$M$       \\ \hline
$\{4,5\}$	& $120$	&	$24$ \\
$\{4,6\}$ &	$24$		&	$4$ \\
$\{8,3\}$	&	$48$		&	$16$ \\
$\{8,4\}$	&	$16$		&	$4$ \\
$\{8,8\}$	&	$8$		&	$1$	\\
$\{10,3\}$	&	$120$	&	$40$	\\
$\{12,3\}$	&	$48$		&	$16$	\\
$\{12,4\}$	&	$24$		&	$6$	\\
$\{14,3\}$	&	$294$	&	$98$	\\
$\{16,3\}$	& $384$	&	$128$	\\
$\{18,3\}$	&	$162$	&	$54$
\end{tabular}
\end{ruledtabular}
\caption{Number of sites $M$ per Bravais unit cell for various hyperbolic lattices, computed according to Eq.~(\ref{MxMyMz-proper}) from the order $|\c{G}^+|$ of the point group tabulated in Ref.~\cite{conder2007}.}
\label{tab:sitespercell}
\end{table}

Several recent studies have explored the spontaneous breakdown of lattice symmetries in half-filled correlated models on $\{p,q\}$ hyperbolic lattices with $p$ even, which are bipartite. At half filling on a bipartite lattice, one expects a gapped ground state with sublattice imbalance of charge or spin, i.e., charge-density wave (CDW) or N\'eel antiferromagnetic (AFM) order~\cite{auerbach}. Bosons with on-site and nearest-neighbor repulsion were studied by quantum Monte Carlo (QMC) on the $\{4,5\}$ and $\{4,6\}$ lattices~\cite{zhu2021}, with insulating CDW ground states found at half filling. Also with QMC, fermionic Hubbard and spin-1/2 Heisenberg models were studied on the $\{10,3\}$, $\{8,3\}$, and $\{8,8\}$ lattices~\cite{gotz2024}, with AFM insulators found at half filling. Mean-field studies of spinless or spinful fermions with repulsive interactions likewise predicted CDW or AFM insulating ground states at half filling on the $\{10,3\}$, $\{14,3\}$, and $\{18,3\}$ lattices~\cite{roy2024,leong2025b,leong2025c}; the $\{12,3\}$ and $\{12,4\}$ lattices~\cite{gluscevich2025}; and the $\{8,4\}$ and $\{16,3\}$ lattices~\cite{leong2025}.

From the LSM perspective developed here, the relevant filling $\nu$ is the number of particles per Bravais unit cell. In Table~\ref{tab:sitespercell}, we list the number of sites per unit cell for each of the lattices listed above. We find that this number is even for all lattices considered, except the $\{8,8\}$ lattice; such that at half filling $\nu$ is integer in all cases except the $\{8,8\}$ lattice, where $\nu=\half$. Thus, only the AFM insulator on the $\{8,8\}$ lattice is \emph{truly} half filled, and also breaks the Fuchsian translation symmetry; all other examples are similar to CDW or AFM states on the honeycomb lattice, which break point-group symmetries but not translation symmetry ($\b{q}=0$ orders).

\subsection{Fractionalized insulators}

Recent studies have constructed exactly soluble spin-1/2 Kitaev models on hyperbolic lattices with fractionalized spin-liquid ground states~\cite{lenggenhager2025,dusel2025,mosseri2025,vidal2025}. However, the question whether spin liquids can exist in hyperbolic lattices more generally---and in particular, \emph{symmetric} spin liquids with unbroken spin-rotation and lattice symmetries~\cite{wen2002}, following Anderson's original paradigm~\cite{anderson1973}---remains open.

\begin{figure}[t]
\includegraphics[width=0.8\columnwidth]{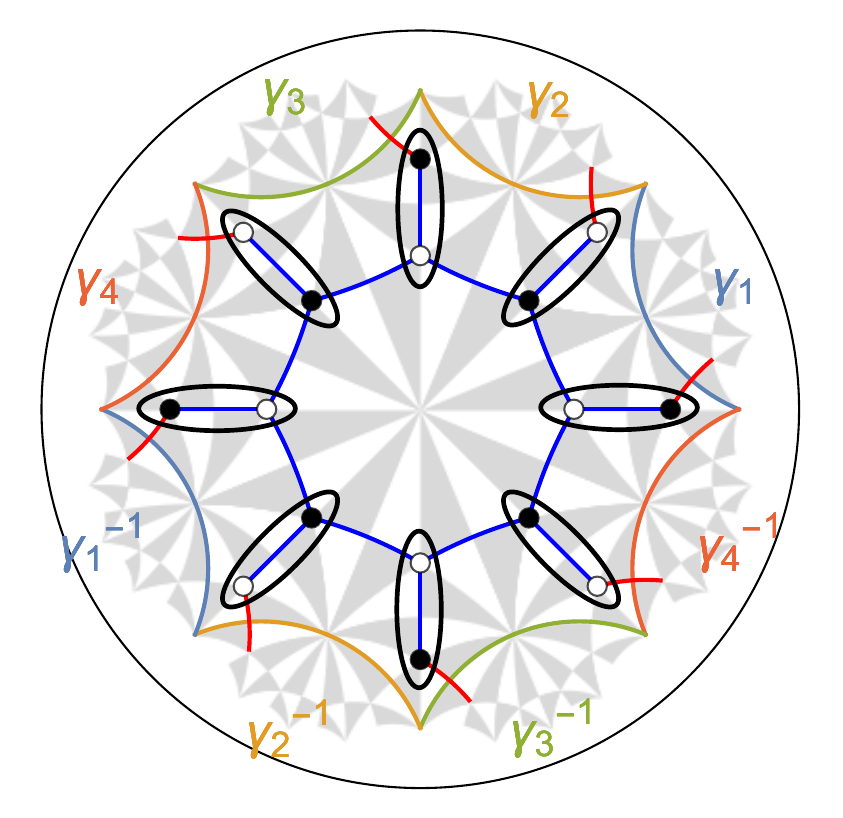}
\caption{Example of spin-1/2 valence-bond solid (VBS) state on the $\{8,3\}$ lattice. A product state of singlet dimers (black ellipses) on nearest-neighbor bonds (blue geodesics) gives a quantum paramagnetic state that preserves spin-rotation symmetry and Fuchsian translation symmetry (Bravais unit cell indicated as an octagon with geodesic edges related by elementary translations $\gamma_1,\ldots,\gamma_4$; red geodesics denote intercell bonds). Such a state could be stabilized by a translation invariant Hamiltonian, e.g., one with strong antiferromagnetic exchange on the bonds hosting the dimers.}
\label{fig:VBS}
\end{figure}

We argue in this section that the LSM constraints derived here give us a clue as to which families of hyperbolic lattices maybe be promising candidates for hosting symmetric spin liquids (or more generally, fractionalized ground states with unbroken symmetries). Consider first a spin-1/2 model on a $\{p,q\}$ lattice with an \emph{even} number of sites per Bravais unit cell, such as the $\{8,3\}$ lattice. Because the filling $\nu$ per unit cell is integer, our LSM result implies that a unique gapped ground state is in principle possible on such lattices, i.e., it should be possible to construct a trivial quantum paramagnet\footnote{We focus here on preserving the spin-rotation and translation symmetries assumed in the LSM theorem. An interesting problem, left for future work, is to construct spin-1/2 quantum paramagnets that are also \emph{featureless}, i.e., that preserve not only translation but also point-group symmetries~\cite{affleck1987,affleck1988b,kimchi2013,ware2015,jian2016,kim2016}. The VBS state depicted in Fig.~\ref{fig:VBS} preserves some, but not all point-group symmetries of the $\{8,3\}$ lattice.}. Indeed, having an even number of sites in the unit cell, one can construct a valence-bond solid (VBS) state as a product state of singlet dimers which preserves the Fuchsian translation symmetry (Fig.~\ref{fig:VBS}). This can be viewed as a hyperbolic analog of the $\b{q}=0$ staggered VBS or ``lattice nematic'' state on the honeycomb lattice~\cite{fouet2001,mulder2010,xu2011}. Such a state can be found as the unique symmetry-preserving ground state of a suitable Hamiltonian\footnote{Here again we ignore point-group symmetries, which are broken spontaneously in Refs.~\cite{fouet2001,mulder2010,xu2011} and lead to additional ground-state degeneracies not accounted for in the ``translation-only'' LSM theorem. By explicitly breaking point-group symmetries at the level of the Hamiltonian while preserving translation symmetry, the ground state can be made unique.}. Turning the argument around, if a non-magnetic, translation-invariant, gapped ground state is found on a $\{p,q\}$ lattice with an even number of spin-1/2 degrees of freedom per Bravais unit cell, there is no guarantee it will be a fractionalized spin liquid.

By contrast, if the number of spin-1/2 degrees of freedom per unit cell is \emph{odd}, as for the $\{8,8\}$ lattice, then $\nu\in\mathbb{Z}+\half$ and our LSM arguments imply that a non-magnetic, translation-invariant, gapped ground state on this lattice must be a fractionalized spin liquid. This suggests that frustrated spin models on such lattices are promising routes to finding hyperbolic analogs of symmetric spin liquids. An example of such model would be an antiferromagnetic Heisenberg model with nearest-neighbor ($J_1$) and next-nearest-neighbor ($J_2$) interactions on the $\{8,8\}$ lattice, as a direct generalization of the $J_1$--$J_2$ Heisenberg antiferromagnet on the square lattice~\cite{read1991,sachdev1991}.

What other regular $\{p,q\}$ lattices possess an odd number $M$ of sites per unit cell? In Euclidean space, besides the square ($\{4,4\}$) lattice, the triangular ($\{3,6\}$) lattice also satisfies this condition, and has accordingly been investigated extensively in the search for spin-liquid ground states~\cite{anderson1973,fazekas1974,sachdev1992,moessner2001b,wang2006,zhu2018}. In hyperbolic space, Ref.~\cite{boettcher2022} presented five infinite families of $\{p,q\}$ lattices (see Table~III therein), and conjectured that three of those families have $M$ even, while the remaining two have $M=1$. The first of those $M=1$ families is the bipartite $\{4g,4g\}$ lattice, which was already discussed in Sec.~\ref{sec:hyp} and generalizes the square lattice. The second is the $\{2g+1,4g+2\}$ lattice, which generalizes the triangular lattice. In Appendix~\ref{app:triangular}, we prove the conjecture posed in Ref.~\cite{boettcher2022} by showing rigorously that the $\{2g+1,4g+2\}$ lattice is Bravais, i.e., its unit cell contains a single site. Since $p=2g+1$ is odd, the $\{2g+1,4g+2\}$ family of lattices is not bipartite and thus can also serve as a promising starting point to explore the consequences of magnetic frustration and to search for symmetric spin-liquid ground states.

Besides gapped spin liquids, another class of fractionalized insulators explored recently in the context of hyperbolic lattices is fractional Chern insulators (FCI)~\cite{he2024,he2025,he2025b}. Original studies of FCI found fractionalized ground states in models of interacting Chern insulators at fractional fillings, e.g., spinless fermions on the checkerboard lattice~\cite{sun2011,neupert2011} at fillings $\nu=\frac{1}{3}$~\cite{neupert2011,sheng2011,regnault2011} and $\nu=\frac{1}{5}$~\cite{sheng2011}; spinless fermions on the honeycomb~\cite{haldane1988} and kagome~\cite{tang2011} lattices at filling $\nu=\frac{1}{3}$~\cite{wu2012}; and hardcore bosons on the honeycomb and checkerboard lattices at fillings $\nu=\frac{1}{2}$ and $\nu=\frac{1}{4}$~\cite{wang2011}. The models considered in those studies preserve translation invariance\footnote{Recall that Chern insulators possess zero magnetic flux per unit cell~\cite{haldane1988}, obviating the need for an (enlarged) magnetic unit cell.} (with PBC) and the $U(1)$ symmetry of particle-number conservation, thus the LSM theorem rules out trivial ground states at the fractional fillings considered. Importantly, in all those studies, the fractional filling $\nu$ is defined as the filling \emph{per Bravais unit cell} (which contains two sites in the case of the honeycomb and checkerboard lattices, and three sites in the case of the kagome lattice). Here, we simply point out that while Refs.~\cite{he2024,he2025} reported numerical evidence for $\nu=\frac{1}{2}$ bosonic FCI states in hyperbolic kagome models, the filling per Bravais unit cell used in those studies does not correspond to $\frac{1}{2}$. The number of sites per Bravais unit cell of the heptagon-kagome (HKG), octagon-kagome (OKG), and nonagon-kagome (NKG) lattices studied in Refs.~\cite{he2024,he2025} is 84, 24, and 162, respectively (see Appendix~\ref{app:wyckoff}). Numerical studies were performed with four or five bosons in 98 sites (HKG), 104 sites (OKG), and 108 sites (NKG), resulting in fillings $\nu$ per unit cell of $\frac{24}{7}$ or $\frac{30}{7}$ (HKG), $\frac{12}{13}$ or $\frac{15}{13}$ (OKG), and $6$ or $\frac{15}{2}$ (NKG). In particular, the LSM constraints presented here imply that a trivial ground state is allowed at filling $\nu=6$ on the NKG lattice.

We conclude this section with a few words on topological degeneracy. In 2D insulating phases with topological order, the presence of anyons implies a ground-state degeneracy on genus-$h$ surfaces that grows exponentially with $h$~\cite{witten1989,moore1989,wen1989,wen1990}. Thus, for such phases, the LSM result (\ref{GSD}) can at most be a lower bound on the true ground-state degeneracy. For example, in a system with $\mathbb{Z}_2$ topological order, the distinct ground state reached upon $2\pi$ flux threading (assuming a suitable aspect ratio of the lattice) is locally indistinguishable from the original state, and thus must correspond to the spontaneous creation of a global $\mathbb{Z}_2$ flux (vison) through a handle of the surface~\cite{paramekanti2004}. There are two possible fluxes per handle, thus the topological degeneracy is $2^{2h}=4^h$. However, as discussed in Sec.~\ref{sec:flux-hyp}, in the hyperbolic case many of those flux configurations are not accessible in a translationally invariant gauge, and thus cannot be reached by the flux threading procedure utilized to derived LSM constraints.

\section{Conclusion}
\label{sec:conclusion}

In this work, we derive general symmetry-enforced constraints on the behavior of quantum many-body systems on hyperbolic lattices. By generalizing Oshikawa's topological argument to hyperbolic lattices, we propose an analog of the Lieb-Schultz-Mattis (LSM) theorem for those lattices. The non-Euclidean geometry of hyperbolic lattices necessitates several modifications to the usual flux threading and momentum counting procedures, which we handle by extending to the many-body realm concepts and methods from the hyperbolic band theory (HBT) of single-particle tight-binding models. In the presence of (non-Euclidean) translation symmetry and a global $U(1)$ symmetry, we show that the particle filling $\nu$ places constraints on the ground-state physics, viz., a trivial ground state is generically allowed only at integer fillings. We derive an explicit lower bound on the ground-state degeneracy and illustrate through numerical computations for the $\{8,8\}$ lattice that rational, non-integer fillings typically forbid such a trivial ground state. A key result of our work is that the LSM filling $\nu$, which constrains the ground-state degeneracy, must be correctly computed as the number of particles \emph{per Bravais unit cell} of the non-Euclidean translation group, which is itself a nontrivial geometric property of the hyperbolic lattice. We analyze from the LSM point of view recent studies of hyperbolic band insulators, charge-density wave (CDW) and antiferromagnetic (AFM) ground states, spin liquids, and fractional Chern insulators (FCI). We propose frustrated spin-1/2 models on the $\{4g,4g\}$ and $\{2g+1,4g+2\}$ families of hyperbolic lattices ($g\geq 2$) as promising candidates for the observation of symmetric spin liquids with an odd number (one) of spins per unit cell. Indeed, for spin-$1/2$ systems at zero magnetization, the equivalent LSM filling $\nu=M/2$ is fractional (half-odd-integer) if and only if the number $M$ of spins per Bravais unit cell is odd. More generally, for a spin-$S$ system, a Bravais unit cell with $M$ spins at magnetization $m$ per spin has filling $\nu=M(S+m)$, and a LSM obstruction to a trivial ground state occurs whenever $M(S+m)\notin\mathbb{Z}$.

Our work opens up a multitude of research directions within the field of hyperbolic quantum many-body physics. As with the original LSM theorem~\cite{paramekanti2004}, we expect our results can be used to provide diagnostic tools to distinguish topological order from symmetry breaking in numerical exact diagonalization studies of quantum many-body Hamiltonians on finite hyperbolic lattices\footnote{We note however in this context that the energy splitting between topologically degenerate ground states on a hyperbolic lattice is generically expected to decay with the total system size as a \emph{power law} rather than exponentially~\cite{mosseri2025}, which can also make topological (quasi-)degeneracy in finite size hard to distinguish from gapless excitations~\cite{bonderson2013}.}. In the search for hyperbolic FCI, it would be interesting to complement the studies of Refs.~\cite{he2024,he2025,he2025b}---where the disk geometry and trapping potential break the Fuchsian translation symmetry explicitly--- with numerical studies on translation-invariant PBC clusters at the appropriate rational LSM filling (e.g., $\nu=\frac{1}{2}$ for hardcore bosons). As proofs of principle, it may also be desirable to theoretically ``engineer'' other fractionalized analogs of hyperbolic Chern insulators more directly~\cite{maciejko2013,zhong2013} so as to analyze their behavior under flux threading explicitly.

An important open problem is to formally extend our results on finite PBC clusters to the thermodynamic limit. For noninteracting Hamiltonians, the bulk spectral properties of infinite hyperbolic lattices are captured by coherent sequences, i.e., nested sequences $\Gamma^{(1)}\vartriangleright\Gamma^{(2)}\vartriangleright\cdots$
of finite-index normal subgroups $\Gamma^{(m)}\nsubg\Gamma$ with trivial total intersection $\bigcap_{m}^\infty\Gamma^{(m)}\!=\!\{e\}$~\cite{lux2024,lux2023,lenggenhager2023}. Assuming that such coherent sequences also capture the thermodynamic limit of interacting Hamiltonians on hyperbolic lattices, it would be interesting to explicitly construct various coherent sequences and verify with the aid of Eq.~(\ref{GSD}) whether nontrivial LSM constraints persist along such sequences.

Another interesting future direction in which to generalize our results is to consider hyperbolic lattices in a magnetic field~\cite{ikeda2021,stegmaier2022,ikeda2023,kong2025}. First, one can consider an applied magnetic field with uniform rational flux $\phi=2\pi p/q$ per unit cell ($p,q$ coprime). In the Euclidean case with e.g. noninteracting electrons, each Bloch band splits into $q$ subbands such that a system at certain fractional fillings (e.g., $\nu=1/q$) can form an insulator with a unique, gapped ground state~\cite{hofstadter1976,thouless1982}. The resolution to this apparent paradox from the LSM perspective is that the magnetic translation operators~\cite{zak1964} $\tilde{T}_1,\tilde{T}_2$ which commute with the Hamiltonian only form a projective representation $\tilde{T}_1\tilde{T}_2\tilde{T}_1^{-1}\tilde{T}_2^{-1}=e^{2\pi ip/q}$ of the translation group $\mathbb{Z}^2$. To obtain a linear representation one must instead use translation operators such as $\tilde{T}_1^q$ and $\tilde{T}_2$ which correspond to a magnetic unit cell $q$ times as large as the original unit cell; thus a fractional filling $\nu=1/q$ of the original unit cell corresponds to integer filling of the magnetic unit cell. Note that those same arguments also apply if the flux is spontaneously generated through many-body interactions~\cite{paramekanti2004}. It would be interesting to generalize the LSM constraints derived here to hyperbolic lattices with rational flux per unit cell by making use of magnetic Fuchsian translation operators~\cite{ikeda2023}. Relatedly, it would be interesting to derive filling-enforced constraints on the quantized Hall response~\cite{sun2024} of hyperbolic lattices by adapting the methods of Ref.~\cite{lu2020} to the hyperbolic setting as done here.

To bring our concluding remarks to a close, we list three additional possible directions: (1) reformulate the LSM constraints derived here in the language of anomalies~\cite{cheng2023,zou2026}, i.e., as a mixed 't Hooft anomaly between Fuchsian translation symmetry and the $U(1)$ symmetry; (2) generalize the present LSM formalism to account for the full space-group symmetries of hyperbolic lattices; and (3) attempt to construct a hyperbolic analog of Luttinger's theorem, for compressible systems. On (2), it would be interesting to see if the framework of lattice homotopy~\cite{po2017,else2020} could be applied to hyperbolic lattices. We have also already mentioned the related problem of constructing spin-1/2 quantum paramagnets that are featureless, i.e., invariant under all space-group operations. On the far more speculative (3), a full generalization of Oshikawa's proof of the Luttinger theorem~\cite{oshikawa2000b} to hyperbolic lattices is likely to be impossible, given the lack of explicit parametrizations of the moduli spaces of higher-dimensional irreps of the Fuchsian translation group, i.e., the non-Abelian Brillouin zones~\cite{maciejko2022,cheng2022,kienzle2022,lenggenhager2023,shankar2024}. However, as a starting point, it may be possible to prove a more modest ``Abelian'' Luttinger's theorem constraining the volume of a ``partial'' Fermi surface in Abelian reciprocal space.

\acknowledgements

We thank I.~Boettcher and C.~Sun for helpful discussions, and T.~Bzdu\v{s}ek and P.~M.~Lenggenhager for collaborations on related projects. J.M. was supported by NSERC Discovery Grants No. RGPIN-2020-06999 and RGPAS-2020-00064 and Quantum Horizons Alberta (QHA).

\appendix
\numberwithin{equation}{section}
\numberwithin{figure}{section}

\section{Rank of sublattice $\Lambda_\PBC\subset\mathbb{Z}^{2g}$}
\label{app:rankLambda}

In this Appendix, we show that $\Lambda_\PBC$ is a full-rank sublattice of $\mathbb{Z}^{2g}$, i.e., it also has rank (dimension) $2g$, and thus possesses $2g$ linearly independent basis vectors. Throughout this and other appendices, we denote subset inclusion by $\subset$, subgroup inclusion by $<$, and normal subgroup inclusion by $\nsubg$. Defining $\Gamma^{(1)}=[\Gamma,\Gamma]$ as the commutator subgroup (or derived subgroup) of $\Gamma$, $\Lambda_\PBC$ is the image of the normal subgroup $\Gamma_\PBC\nsubg\Gamma$ under the Hurewicz homomorphism (\ref{Lambda}), which is the quotient homomorphism by $\Gamma^{(1)}$,
\begin{align}
\Lambda:\Gamma\rightarrow\Gamma/\Gamma^{(1)}\cong\mathbb{Z}^{2g}.
\end{align}
According to the first and second isomorphism theorems~\cite{Robinson}, the image of a subgroup $H<\Gamma$ under the quotient homomorphism $f:\Gamma\rightarrow\Gamma/N$ where $N\nsubg\Gamma$ is $f(H)=NH/N$, where the product $NH$ is
\begin{align}
NH=\{nh:n\in N,h\in H\}=\{Nh:h\in H\}.
\end{align}
Note that $N$ and $H$ permute, $NH=HN$, and thus $NH<\Gamma$. Additionally, $N\nsubg NH$, and thus $NH/N$ is a group. Assuming $M$ cosets, $NH/N$ can be written as
\begin{align}\label{NHmodN}
NH/N=\{N,Nh_2,\ldots,Nh_M\},
\end{align}
where $h_1=e,h_2,\ldots,h_M$ are coset representatives, with $e$ the identity element. Choosing $N=\Gamma^{(1)}$, we thus obtain
\begin{align}
\Lambda_\PBC=\Gamma^{(1)}\Gamma_\PBC/\Gamma^{(1)}.
\end{align}
Since $\Lambda_\PBC$ is a subgroup of the Abelian group $\mathbb{Z}^{2g}$, it is necessarily also Abelian and normal in $\mathbb{Z}^{2g}$, thus we can form the quotient of the two lattices
\begin{align}\label{ZoverL}
\mathbb{Z}^{2g}/\Lambda_\PBC&=(\Gamma/\Gamma^{(1)})/(\Gamma^{(1)}\Gamma_\PBC/\Gamma^{(1)})\nn\\
&\cong\Gamma/(\Gamma^{(1)}\Gamma_\PBC),
\end{align}
where we have used the third isomorphism theorem~\cite{Robinson}. Note that the theorem applies since $\Gamma^{(1)}$ is obviously a subgroup of $\Gamma^{(1)}\Gamma_\PBC$, and both $\Gamma^{(1)}$ and $\Gamma^{(1)}\Gamma_\PBC$ are normal in $\Gamma$. The latter property holds because both $\Gamma^{(1)}$ and $\Gamma_\PBC$ are normal in $\Gamma$, and thus for any $\gamma\in\Gamma$, $\gamma\Gamma^{(1)}\Gamma_\PBC\gamma^{-1}=(\gamma\Gamma^{(1)}\gamma^{-1})(\gamma\Gamma_\PBC\gamma^{-1})=\Gamma^{(1)}\Gamma_\PBC$.

We next show that the quotient (\ref{ZoverL}) is isomorphic to the abelianization of the factor group $G\equiv\Gamma/\Gamma_\PBC$:
\begin{align}\label{abelianizationGG1}
\mathbb{Z}^{2g}/\Lambda_\PBC\cong G_\text{ab}=G/G^{(1)},
\end{align}
with $G^{(1)}=[G,G]$. Instead of $\Lambda$, consider now the quotient homomorphism $\phi:\Gamma\rightarrow\Gamma/\Gamma_\PBC\cong G$. The image of $\Gamma^{(1)}$ under this homomorphism is $\phi(\Gamma^{(1)})=\Gamma_\PBC\Gamma^{(1)}/\Gamma_\PBC$. The third isomorphism theorem can be applied again with $\Gamma^{(1)}$ and $\Gamma_\PBC$ interchanged, because both $\Gamma_\PBC$ and $\Gamma_\PBC\Gamma^{(1)}=\Gamma^{(1)}\Gamma_\PBC$ are normal in $\Gamma$, and $\Gamma_\PBC$ is obviously a subgroup of $\Gamma_\PBC\Gamma^{(1)}$. Thus, we can compute the quotient
\begin{align}
G/\phi(\Gamma^{(1)})&=(\Gamma/\Gamma_\PBC)/(\Gamma_\PBC\Gamma^{(1)}/\Gamma_\PBC)\nn\\
&\cong\Gamma/(\Gamma_\PBC\Gamma^{(1)})\nn\\
&\cong\mathbb{Z}^{2g}/\Lambda_\PBC,
\end{align}
using Eq.~(\ref{ZoverL}) and the fact that $\Gamma_\PBC$ and $\Gamma^{(1)}$ permute. On the one hand, using Eq.~(\ref{NHmodN}), the image $\phi(\Gamma^{(1)})$ is the set of all distinct elements of the form $\Gamma_\PBC\eta$, where $\eta$ is an element of the commutator subgroup $\Gamma^{(1)}$, i.e., $\eta=[\gamma_1,\gamma_1']\cdots[\gamma_n,\gamma_n']$ for some $\gamma_i,\gamma_i'\in\Gamma$, $i=1,\ldots,n$, and the commutator of two elements is $[\gamma,\gamma']\equiv\gamma\gamma'\gamma^{-1}\gamma^{\prime-1}$. On the other hand, the elements of $G$ are the distinct cosets $\Gamma_\PBC\gamma$ with $\gamma\in\Gamma$, such that the commutator subgroup $G^{(1)}$ is the set of distinct elements of the form $[\Gamma_\PBC\gamma_1,\Gamma_\PBC\gamma_1']\cdots [\Gamma_\PBC\gamma_n,\Gamma_\PBC\gamma_n']=\Gamma_\PBC[\gamma_1,\gamma_1']\cdots[\gamma_n,\gamma_n']$, where we have used the normality of $\Gamma_\PBC$. Thus $\phi(\Gamma^{(1)})\cong G^{(1)}$, and we have proved Eq.~(\ref{abelianizationGG1}).

The index $I_\PBC\equiv[\mathbb{Z}^{2g}:\Lambda_\PBC]$ of the sublattice $\Lambda_\PBC$ in $\mathbb{Z}^{2g}$ is the number of points of $\mathbb{Z}^{2g}$ inside any unit cell of $\Lambda_\PBC$. It is also the index of $\Lambda_\PBC$ as a subgroup of the Abelian group $\mathbb{Z}^{2g}$, thus $I_\PBC=|G/G^{(1)}|$. Recall that the total number of sites $V=[\Gamma:\Gamma_\PBC]=|G|$ of the hyperbolic lattice is finite by assumption, thus by Lagrange's theorem, the index
\begin{align}\label{IPBC}
I_\PBC=|G/G^{(1)}|=V/|G^{(1)}|
\end{align}
is also finite, where $|G^{(1)}|$ divides $V$. Since a unit cell of $\Lambda_\PBC$ only contains finitely many points of $\mathbb{Z}^{2g}$, $\Lambda_\PBC$ extends infinitely in all $2g$ directions and has rank $2g$.

\section{Number of sites per Bravais unit cell}
\label{app:wyckoff}

In this Appendix, we derive formulas [Eqs.~(\ref{MxMyMz}) and (\ref{MxMyMz-proper})] for the number of sites per Bravais unit cell for three standard types of hyperbolic lattices~\cite{HyperCells,HyperBloch}. The space group of the $\{p,q\}$ lattice with $(p-2)(q-2)>4$ is the hyperbolic triangle group $\Delta\equiv\Delta(2,q,p)$, an infinite discrete group defined by the presentation~\cite{boettcher2022}
\begin{align}\label{FullTriangleGroup}
\Delta=\langle a,b,c|a^2,b^2,c^2,(ab)^2,(bc)^q,(ca)^p\rangle,
\end{align}
where $a,b,c$ are reflections in the sides of a hyperbolic triangle with interior rangles $\frac{\pi}{2}$, $\frac{\pi}{q}$, $\frac{\pi}{p}$, corresponding respectively to three corners $x,y,z$. Those three corners correspond to three \emph{Wyckoff positions}, i.e., positions $s\in\{x,y,z\}$ in the hyperbolic plane with nontrivial site-symmetry group or stabilizer subgroup $\Delta_s<\Delta$~\cite{lenggenhager2023}. For each corner $x,y,z$, the respective stabilizer group is generated by the two hyperbolic reflections in the sides joined at this corner:
\begin{align}
\Delta_x&=\langle a,b|a^2,b^2,(ab)^2\rangle\cong D_2,\\
\Delta_y&=\langle b,c|b^2,c^2,(bc)^q\rangle\cong D_q,\\
\Delta_z&=\langle c,a|c^2,a^2,(ca)^p\rangle\cong D_p.
\end{align}
These are Coxeter groups~\cite{CoxeterPolytopes} isomorphic to a dihedral group $D_n$ of order $2n$, thus $|\Delta_x|=4$, $|\Delta_y|=2q$, and $|\Delta_z|=2p$. For given $p$ and $q$, three standard lattices can be defined by keeping only:
\begin{itemize}
\item $y$-type vertices (the $\{p,q\}$ lattice itself);
\item $z$-type vertices (the dual $\{q,p\}$ lattice); or
\item $x$-type vertices (e.g., the line graph of the $\{p,q\}$ lattice, which for $q=3$ corresponds to a kagome-like lattice of corner-sharing triangles~\cite{kollar2020}).
\end{itemize}
We wish to derive formulas expressing the number $M_x,M_y,M_z$ of vertices of each type inside the Bravais unit cell associated with the Fuchsian translation group $\Gamma\nsubg\Delta$, as functions of the order $|\c{G}|$ of the point group $\c{G}=\Delta/\Gamma$, because such point groups and their orders have been tabulated for a wide range of $\{p,q\}$ lattices~\cite{conder2007,HyperCells,HyperBloch}.

Given a reference site $s_0$ of type $s$, the set of all such sites in the infinite hyperbolic plane is given by the orbit $s_0\cdot\Delta$, where we define a right group action $s\cdot g$ of $\Delta\ni g$ on points $s$ in the hyperbolic plane. By the orbit-stabilizer theorem, the points in this orbit are in one-to-one correspondence with the elements of the right coset space $\Delta_s\backslash\Delta$. Thus, the sites of type $s$ within the infinite lattice can be labeled as right cosets in $\Delta_s\backslash\Delta$.

To count how many such sites are contained in a single Bravais unit cell, first observe that two points $s,s'$ correspond to the same site in the unit cell if and only if they are related by a translation $\gamma\in\Gamma$, i.e., $s'=s\cdot\gamma$. Writing $s=s_0\cdot g$ and $s'=s_0\cdot g'$ with $g,g'\in\Delta$, we have $s_0\cdot g'=s_0\cdot(g\gamma)$, which implies there exists an element $h\in\Delta_s$ of the stabilizer group such that $g'=hg\gamma$. This defines an equivalence relation $g\sim g'$ on $\Delta$. Given two subgroups $\Delta_s$ and $\Gamma$ of $\Delta$, the equivalence class $\Delta_s g\Gamma$ of $g$ under this equivalence relation is called the $(\Delta_s,\Gamma)$-\emph{double coset}~\cite{Robinson} of $g\in\Delta$, and the set of all $(\Delta_s,\Gamma)$-double cosets is denoted $\Delta_s\backslash\Delta/\Gamma$ (``$\Delta$ mod $\Gamma$ mod $\Delta_s$''). Thus, the number $M_s$ of sites of type $s$ within the Bravais unit cell is equal to the number of double cosets in $\Delta_s\backslash\Delta/\Gamma$.

To relate $M_s$ to $|\c{G}|=[\Delta:\Gamma]$, we use the following formula, proved immediately after Eq.~(\ref{Ms}):
\begin{align}\label{DoubleCosetEq}
[\Delta:\Gamma]=\sum_{\Delta_sg\Gamma\in\Delta_s\backslash\Delta/\Gamma}[\Delta_s:\Delta_s\cap g\Gamma g^{-1}],
\end{align}
where the sum is over double-coset representatives $g$. First, we prove that $\Delta_s\cap g\Gamma g^{-1}=\{e\}$, the trivial group. All elements in $\Delta_s$ have finite order since $\Delta_s$ is a finite group. By contrast, since $\Gamma$ is torsion free, all its nontrivial elements have infinite order. Since the order of an element is preserved under conjugation, all nontrivial elements of $g\Gamma g^{-1}$ also have infinite order. Thus, $\Delta_s\cap g\Gamma g^{-1}=\{e\}$ and we obtain
\begin{align}\label{Ms}
M_s=|\Delta_s\backslash\Delta/\Gamma|=\frac{[\Delta:\Gamma]}{|\Delta_s|}=\frac{|\c{G}|}{|\Delta_s|},
\end{align}
where $|\c{G}|$ is the order of the point group. Thus, we obtain the main results of this Appendix:
\begin{align}\label{MxMyMz}
M_x=\frac{|\c{G}|/2}{2},\hspace{5mm}M_y=\frac{|\c{G}|/2}{q},\hspace{5mm}M_z=\frac{|\c{G}|/2}{p},
\end{align}
which respectively give the number of sites of type $x$, $y$, or $z$ inside the Bravais unit cell.

To prove Eq.~(\ref{DoubleCosetEq}), consider the space $\Delta/\Gamma$ of left $\Gamma$-cosets of $\Delta$. For a given left $\Gamma$-coset $[g]\equiv g\Gamma$, the double coset $\Delta_s g\Gamma=\Delta_s[g]$ can also be viewed as the orbit of $[g]$ under a left action of $\Delta_s$ by multiplication $h\cdot [g]=(hg)\Gamma=[hg]$, $h\in\Delta_s$. Once again invoking the orbit-stabilizer theorem, the number of left $\Gamma$-cosets in this orbit is $[\Delta_s:\Stab_{\Delta_s}([g])]$, where we denote by $\Stab_{\Delta_s}([g])$ the stabilizer of $[g]$ in $\Delta_s$,
\begin{align}\label{Stab}
\Stab_{\Delta_s}([g])=\Delta_s\cap g\Gamma g^{-1}.
\end{align}
Indeed, we have $g\Gamma g^{-1}\cdot[g]=g\Gamma g^{-1}g\Gamma=g\Gamma$, but we need the stabilizer in $\Delta_s$ and thus the intersection with $\Delta_s$ must be taken. Just like ordinary coset decompositions, the double-coset decomposition is a disjoint decomposition of $\Delta$, thus each $\Gamma$-coset belongs to a distinct double coset $\Delta_s g\Gamma$. Therefore, the total number $[\Delta:\Gamma]$ of $\Gamma$-cosets in $\Delta$ is given by summing $[\Delta_s:\Stab_{\Delta_s}([g])]$ over all distinct double cosets $\Delta_s g\Gamma$, where $g$ denotes a choice of double coset representative:
\begin{align}
[\Delta:\Gamma]=\sum_{\Delta_s g\Gamma\in\Delta_s\backslash \Delta/\Gamma}[\Delta_s:\Stab_{\Delta_s}([g])],
\end{align}
from which Eq.~(\ref{DoubleCosetEq}) follows after using Eq.~(\ref{Stab}).

As an illustration of these results, consider the $\{8,3\}$ lattice, whose point group $\c{G}=\Delta(2,3,8)/\Gamma=\mathrm{GL}(2,\mathbb{Z}_3)\rtimes\mathbb{Z}_2$ has $|\c{G}|=96$ elements~\cite{maciejko2021,chen2023}. The Bravais unit cell contains $M_x=(96/2)/2=24$ sites of the octagon-kagome lattice~\cite{bzdusek2022,mosseri2022}, $M_y=(96/2)/3=16$ sites of the $\{8,3\}$ lattice~\cite{boettcher2022}, and $M_z=(96/2)/8=6$ sites of the $\{3,8\}$ lattice. Those formulas can also be used for hyperbolic analogs of the dice lattice~\cite{sutherland1986}, where both $y$ and $z$ sites are kept, giving a Bravais unit cell with $M=M_y+M_z$ sites. For example, the octagon-dice lattice~\cite{bzdusek2022,mosseri2022} contains $M=16+6=22$ sites per unit cell.

Finally, we observe that the groups tabulated in the online database of Ref.~\cite{conder2007} correspond to the \emph{proper} point group $\c{G}^+$, which is the orientation-preserving subgroup of index 2 in the full point group $\c{G}$. As a result, $|\c{G}^+|=|\c{G}|/2$, and Eq.~(\ref{MxMyMz}) can also be written as
\begin{align}\label{MxMyMz-proper}
M_x=\frac{|\c{G}^+|}{2},\hspace{5mm}M_y=\frac{|\c{G}^+|}{q},\hspace{5mm}M_z=\frac{|\c{G}^+|}{p},
\end{align}
where the data for $|\c{G}^+|$ can be directly read off from the ``group order'' entry in the database of Ref.~\cite{conder2007}; see Table~\ref{tab:sitespercell} for a few examples.

\section{The $\{2g+1,4g+2\}$ lattice is Bravais}
\label{app:triangular}

In this Appendix, we prove that the Bravais unit cell of the $\{2g+1,4g+2\}$ lattice contains a single site for any $g\geq 2$, as conjectured in Ref.~\cite{boettcher2022}. The strategy of the proof is as follows. Defining $p=4g+2$ and $q=2g+1$, the number $M_z$ of vertices of the $\{q,p\}$ lattice is given by $M_z=|\c{G}^+|/p$ [see Eq.~(\ref{MxMyMz-proper})], and we wish to show that $M_z=1$. We will prove that $\c{G}^+\cong\mathbb{Z}_{4g+2}$, such that $M_z=|\mathbb{Z}_{4g+2}|/(4g+2)=1$.

The proper point group $\c{G}^+$ is the quotient $\Delta^+/\Gamma$ where $\Gamma$ is normal and torsion free. Here, we define the proper triangle group~\cite{boettcher2022} $\Delta^+\equiv\Delta^+(2,q,p)$ as the orientation-preserving subgroup of index 2 in the full triangle group (\ref{FullTriangleGroup}):
\begin{align}
\Delta^+=\langle x,y,z|x^2,y^q,z^p,xyz\rangle,
\end{align}
where $x=ab$, $y=bc$, $z=ca$ respectively describe rotations of order 2, $q$, $p$ around the corners $x$, $y$, $z$ of the hyperbolic triangle mentioned in Appendix~\ref{app:wyckoff}. We can eliminate one generator ($z$) and one relation ($xyz=e$) and equivalently write
\begin{align}
\Delta^+=\langle x,y|x^2,y^q,(xy)^p\rangle.
\end{align}
If we can construct a surjective group homomorphism
\begin{align}\label{homo}
\phi:\Delta^+\rightarrow\mathbb{Z}_{4g+2},
\end{align}
such that its kernel $\Gamma\equiv\ker\phi\nsubg\Delta^+$ is torsion free, then by the first isomorphism theorem~\cite{Robinson}, $\c{G}^+\equiv\Delta^+/\ker\phi\cong\im\phi=\mathbb{Z}_{4g+2}$. 

First, we show that $\ker\phi$ is torsion free if and only if the homomorphism $\phi$ preserves the order of the elements $x$, $y$, and $xy$, viz.
\begin{align}\label{OrdersHomo}
|\phi(x)|=2,\hspace{5mm}|\phi(y)|=q,\hspace{5mm}|\phi(xy)|=p,
\end{align}
denoting by $|h|$ the order of a group element $h$, i.e., the smallest positive integer $|h|$ such that $h^{|h|}=e$. Note that for any element $h$, $|\phi(h)|$ divides $|h|$ since $\phi(h)^{|h|}=\phi(h^{|h|})=e$. Let $s\in\{x,y,xy\}$ and suppose that $|\phi(s)|$ strictly divides $|s|$, i.e., $\ell=|s|/|\phi(s)|>1$. But then $s^{|\phi(s)|}$ has order $\ell>1$ and belongs to the kernel, since $\phi(s^{|\phi(s)|})=\phi(s)^{|\phi(s)|}=e$. Thus it is necessary for $\ker\phi$ to be torsion free that $|\phi(s)|=|s|$. It is also sufficient, for the following reason. By Theorem 2.10 of Ref.~\cite{Magnus}, every nontrivial torsion element $\gamma$ of $\Delta^+$ is conjugate to some nontrivial power $k$ of $s\in\{x,y,xy\}$, i.e., $\gamma=hs^kh^{-1}$ with $h\in\Delta^+$ and $s^k\neq e$. But since $|\phi(s)|=|s|$ and conjugation preserves the order, $\phi$ maps $\gamma$ to a nontrivial torsion element of $\im\phi$, and thus $\gamma$ cannot be in the kernel.

We represent the cyclic group $\mathbb{Z}_{4g+2}$ as the additive group of integers modulo $4g+2$ and define a homomorphism $\phi:\Delta^+\rightarrow\mathbb{Z}_{4g+2}$ via its action on the generators $x,y$:
\begin{align}\label{homo-gens}
\phi(x)=2g+1,\hspace{5mm}\phi(y)=2.
\end{align}
Since $\mathbb{Z}_{4g+2}$ is Abelian, we have $\phi(xy)=\phi(x)+\phi(y)=2g+3$. We now show that the orders $|\phi(s)|$ for $s\in\{x,y,xy\}$ under this homomorphism are indeed given by Eq.~(\ref{OrdersHomo}). In modular arithmetic, the order of an element $\phi(s)\in\mathbb{Z}_{4g+2}$ is given by the number $b$ of distinct rational points $\phi(s)/(4g+2)$ on the real axis modulo 1. (See the discussion above Eq.~(\ref{GSD}) where the exact same mathematics is at play.) To calculate this number, we simplify the fraction $\phi(s)/(4g+2)=a/b$ such that $a,b$ are coprime, thus $b=(4g+2)/\gcd(\phi(s),4g+2)$. From Eq.~(\ref{homo-gens}), we obtain:
\begin{align}
|\phi(x)|&=\frac{4g+2}{\gcd(2g+1,4g+2)}=\frac{4g+2}{2g+1}=2,\\
|\phi(y)|&=\frac{4g+2}{\gcd(2,4g+2)}=\frac{4g+2}{2}=2g+1,
\end{align}
thus verifying the first two equalities in Eq.~(\ref{OrdersHomo}). Using $\gcd(2g+3,4g+2)=\gcd(2g+3,4)=1$, since $2g+3$ is odd and thus has no common factors with $4$, we verify the third equality in Eq.~(\ref{OrdersHomo}):
\begin{align}
|\phi(xy)|=\frac{4g+2}{\gcd(2g+3,4g+2)}=4g+2.
\end{align}
Thus, we have proved that the homomorphism (\ref{homo-gens}) has a torsion-free kernel. Finally, since $\phi(xy)$ has order exactly $4g+2$, it is a generator of $\mathbb{Z}_{4g+2}$ and thus the homomorphism (\ref{homo-gens}) is surjective, implying our desired conclusion.

\bibliography{hyplsm}

\end{document}